\documentclass[onecolumn,superscriptaddress,prb]{revtex4}
\usepackage{amsmath}
\usepackage{graphicx}
\usepackage{xcolor} 
\usepackage[normalem]{ulem} 

\newcommand{\md}{\mathrm d}

\begin{document}

\title{Optimal Control in Stochastic Thermodynamics}

\author{Steven Blaber}
\email{sblaber@sfu.ca}
\affiliation{Dept.~of Physics, Simon Fraser University, Burnaby, British Columbia V5A 1S6, Canada}
\author{David A.\ Sivak}
\email{dsivak@sfu.ca}
\affiliation{Dept.~of Physics, Simon Fraser University, Burnaby, British Columbia V5A 1S6, Canada}

\begin{abstract}
We review recent progress in optimal control in stochastic thermodynamics. Theoretical advances provide in-depth insight into minimum-dissipation control with either full or limited (parametric) control, and spanning the limits from slow to fast driving and from weak to strong driving. Known exact solutions give a window into the properties of minimum-dissipation control, which are reproduced by approximate methods in the relevant limits. Connections between optimal-transport theory and minimum-dissipation protocols under full control give deep insight into the properties of optimal control and place bounds on the dissipation of thermodynamic processes. Since minimum-dissipation protocols are relatively well understood and advanced approximation methods and numerical techniques for estimating minimum-dissipation protocols have been developed, now is an opportune time for application to chemical and biological systems.
\end{abstract}

\maketitle

\section{Introduction}\label{Introduction}
In this article we review recent progress in optimal control of stochastic thermodynamic systems. We focus on classical isothermal stochastic thermodynamics describing control by linear-response, thermodynamic geometry, and optimal-transport theory.

Historically, modern thermodynamic control began with the study of finite-time thermodynamics of macroscopic systems,~\cite{andresen1977,salamon1977,andresen2011} the natural extension beyond quasistatic (infinitely slow) processes. Fundamentally, any finite-time thermodynamic control will induce some degree of irreversibility, manifesting as energy dissipated into the environment. A goal of finite-time thermodynamics is to quantify and minimize this dissipation through the use of designed control strategies. For example, Ref.~\onlinecite{band1982} studied the optimal cycle for finite-time operation of a heat engine and found that instantaneous jumps in control parameters are necessary to minimize dissipation.

In parallel, a thermodynamic-geometry framework was developed to provide a novel means to describe thermodynamic processes on a smooth (generally Riemannian) manifold.\cite{Weinhold1975,ruppeiner1979,Crooks2007} Ref.~\onlinecite{salamon1983} showed the connections between thermodynamic geometry and minimum-dissipation protocols, opening the door for the development of a geometric description of minimum-dissipation protocols. Although theoretically compelling, the utility of the framework was not fully realized until the development of stochastic thermodynamics.

The aforementioned descriptions focused on macroscopic systems that equilibrate rapidly and whose fluctuations are relatively small. The advent of modern experimental techniques, including single-molecule biophysical experiments, created demand for a theoretical description of the energetics of microscopic systems. Optical tweezers and magnetic traps~\cite{ashkin1970,bustamante2021,polimeno2018,moffitt2008} allow for the precise manipulation of individual polymer strands~\cite{liphardt2002,collin2005,bustamante2000,bustamante2003,woodside2006,neupane2017} and nanoscale molecular machines~\cite{unksov2022} (e.g., ATP synthase,\cite{Toyabe2011,toyabe2012,kawaguchi2014} kinesin,\cite{svoboda1993,svoboda1994,kojima1997,hunt1994} and myosin\cite{greenberg2016,Laakso2008,norstrom2010,nagy2013}). Due to their small scale, these systems are not accurately described by macroscopic thermodynamics since the fluctuations (of order $k_{\rm B}T$) are comparable to the systems' internal energy scales, and the operating speeds of single-molecule experiments and molecular machines are comparable to those systems' natural relaxation times.

To describe these small-scale systems, the field of stochastic thermodynamics was developed, which aims to describe the nonequilibrium energetics of stochastic (fluctuating) microscopic systems~\cite{Seifert2012,Jarzynski2011}. Just like its macroscopic counterpart, a central goal of stochastic thermodynamics is the description of optimal control strategies: methods for performing a given task at minimum energetic cost~\cite{Brown2017,Brown2019}.

\begin{figure}
	\includegraphics[width=\linewidth]{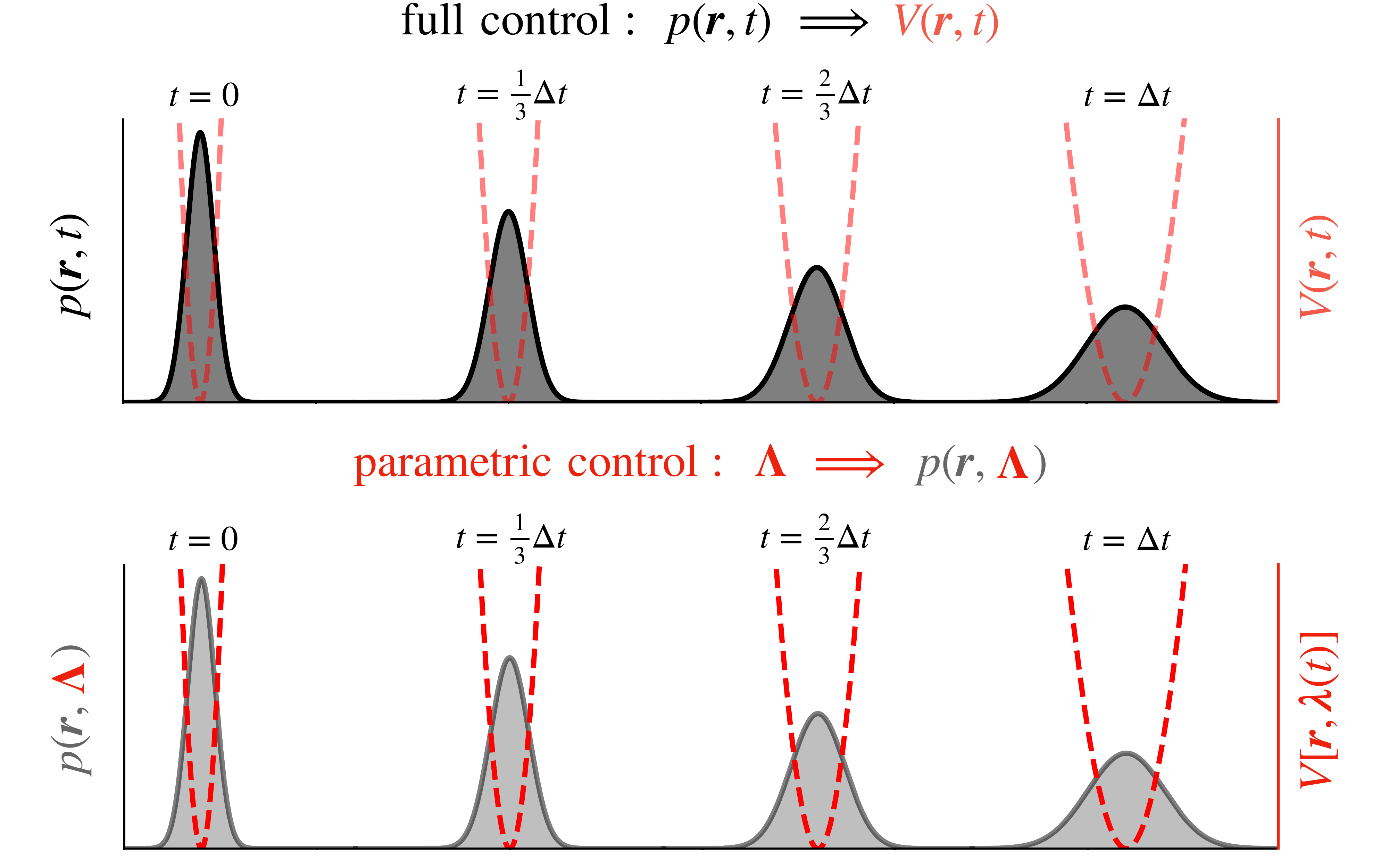}
	\caption{Comparing full control and parametric control. Full control (top) assumes complete control of the probability distribution $p(\boldsymbol{r},t)$ (shaded) which can be optimally driven between the endpoints by a potential $V(\boldsymbol{r},t)$ (red dashed curves). Parametric control (bottom) adjusts a finite number of control parameters $\boldsymbol{\lambda}(t)$ according to a protocol $\boldsymbol{\Lambda}$ between specified endpoints, thereby driving the probability distribution $p(\boldsymbol{r},\boldsymbol{\Lambda})$.}
	\label{Full_vs_parametric}
\end{figure}

There are two distinct but related types of control we consider in this review: full control (section~\ref{Full control})and parametric control (section~\ref{Parametric control}). Full control assumes we have complete control of the probability distribution (Fig.~\ref{Full_vs_parametric}, top). Parametric control adjusts a finite number of control parameters (Fig.~\ref{Full_vs_parametric}, bottom), and in doing so drives the probability distribution.

For either full or parametric control, the exact minimum-dissipation protocol is known if the probability distribution is Gaussian~\cite{Schmiedl2007,abiuso2022,Blaber2022_strong}. These exact solutions provide a glimpse into the properties of optimal control processes. For example, just like the finite-time thermodynamic control described previously, the minimum-dissipation protocol has discontinuous changes in control parameters at the start and end of the protocol but remains continuous between these endpoints~\cite{Schmiedl2007}. These discontinuities are present even for underdamped dynamics~\cite{Gomez2008}. The control-parameter jumps have been observed in a number of different systems~\cite{Gomez2008,Then2008,Esposito2010} and are now well understood and have been shown to be a general feature~\cite{Blaber2021}.

For more general solutions under full control, the study of minimum-dissipation protocols can be mapped onto a problem of optimal-transport theory, a well developed branch of mathematics for which there exist numerous algorithms and methods for determining the optimal-transport map~\cite{villani2009,santambrogio2015}. The connection between minimum-dissipation protocols and optimal-transport theory was first shown in Ref.~\onlinecite{Aurell2011} for overdamped dynamics: the protocol that minimizes dissipation when driving a system obeying overdamped Fokker-Planck dynamics between specified initial and final distributions is governed by the Wasserstein distance~\cite{Zhang2019,nakazato2021,dechant2022,miangolarra2022} and the Benamou-Brenier formula~\cite{benamou2000}. This technique eventually led to new fundamental lower bounds on the average work required for finite-time information erasure~\cite{Proesmans2020,proesmans2020optimal}. Initially only applicable to overdamped dynamics, the connections between optimal-transport theory and minimum-dissipation protocols have recently been shown for discrete-state and quantum systems~\cite{dechant2022,dechant2022geometric,yoshimura2022,zhong2022,van2022,van2022Topological}. 

\begin{figure}
	\includegraphics[width=0.75\linewidth]{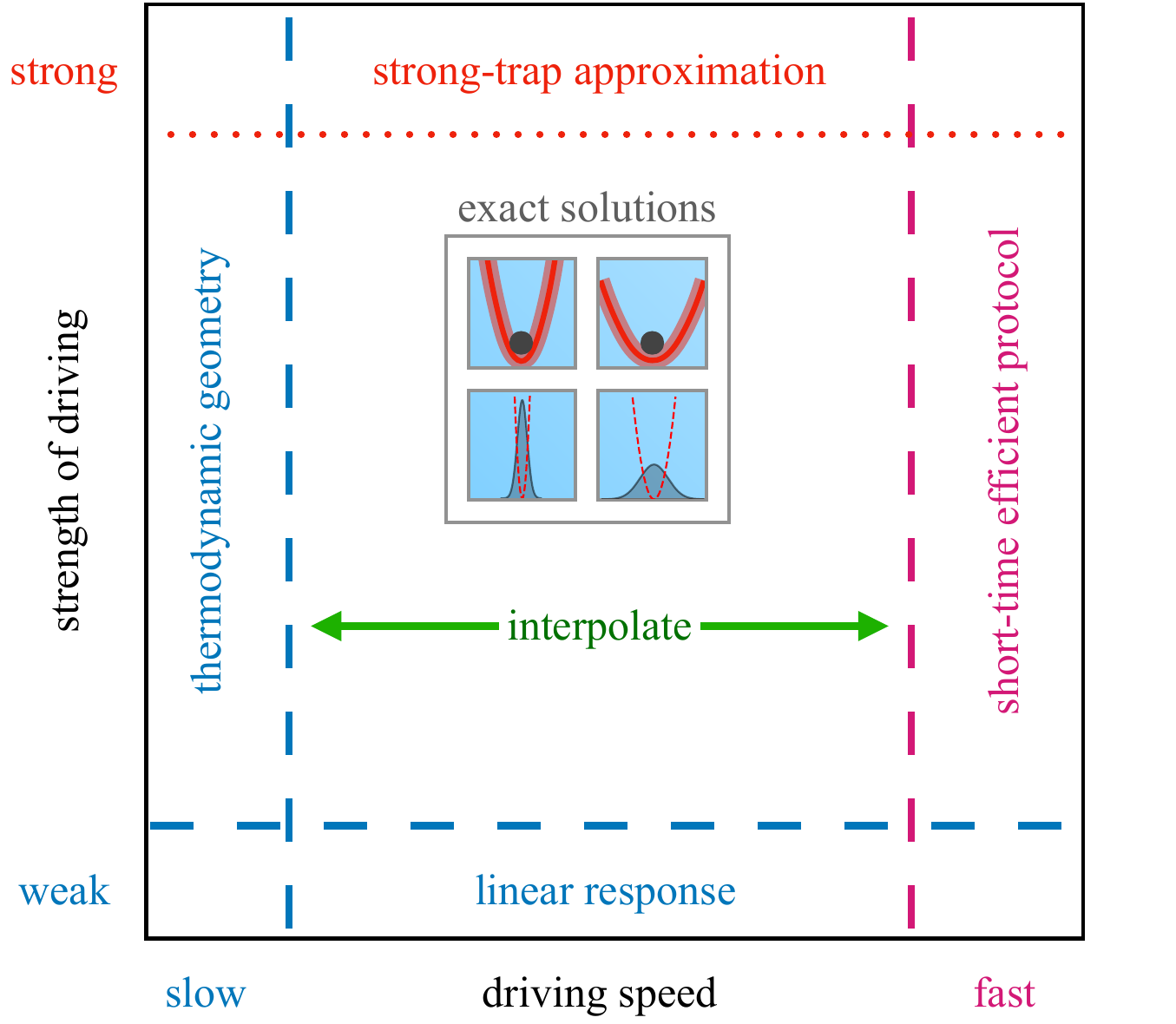}
	\caption{The space of thermodynamic control. Horizontal axis is the driving speed from slow to fast, and the vertical axis is the strength of driving from weak to strong. Linear-response theory is applicable to weak and slow driving (blue), and can be simplified to a thermodynamic-geometry framework for slow driving. Short-time efficient protocols are valid for fast driving (purple) and can be combined with thermodynamic geometry to bridge the space between slow and fast with interpolated protocols (green). The strong-trap approximation (red) is only valid for overdamped dynamics, with region of applicability schematically indicated by a distinct dotted line. Exact solutions for Gaussian distributions serve as a window into the properties of minimum-dissipation protocols and are valid at any driving speed or strength of driving.} 
	\label{Parametric_diagram}
\end{figure}

General solutions for parametric control are typically difficult to determine, although recent progress has been made towards exact solutions for general systems building off of optimal-transport~\cite{zhong2022} or advanced numerical techniques~\cite{Then2008,gingrich2016,engel2022}. Although exact solutions are convenient where possible, the determination of minimum-dissipation protocols can be considerably simplified through approximate methods. Inspired by a diagram presented in Ref.~\onlinecite{bonanca2018}, we schematically show in Fig.~\ref{Parametric_diagram} the limits where minimum-dissipation protocols are known.

Linear-response theory can be used to determine the minimum-dissipation protocol for weak perturbations and performs relatively well at any driving speed and beyond its strict range of validity~\cite{kamizaki2022,bonanca2018}. For slow control, the thermodynamic-geometry framework has been generalized to stochastic thermodynamic systems~\cite{Crooks2007,Sivak2016} and has been used to explore a diverse set of model systems~\cite{Sivak2016,Blaber2020,deffner2020,zulkowski2012,bonancca2014,zulkowski2015,zulkowski2015Quantum,Large2019,Lucero2019,Rotskoff2015,Rotskoff2017,louwerse2022,frim2022,frim2021}, including DNA-pulling experiments~\cite{Tafoya2019} and free-energy estimation~\cite{Blaber2020Skewed}. In the opposite limit of fast control, minimum-dissipation protocols are describe by short-time efficient protocols~\cite{Blaber2021}, which can be combined with the thermodynamic-geometry framework to design interpolated protocols that perform well at any driving speed~\cite{Blaber2022}. Leveraging known solutions from optimal-transport theory, strong control can be described by the strong-trap approximation, yielding explicit solutions for minimum-dissipation protocols~\cite{Blaber2022_strong}.

There are a number of related topics which are not covered in this review, such as optimal control of heat engines{\cite{holubec2021}} (including optimal cycles~\cite{ma2018,zhang2020,abiuso2020,frim2021,frim2022,chen2022} and efficiency at maximum power~\cite{curzon1975,van2005,schmiedl2008efficiency,esposito2009,esposito2010efficiency,brandner2015,proesmans2016,shiraishi2016,ma2018universal,ma2020,miller2020,brandner2020,miangolarra2021,miangolarra2022,watanabe2022}) and optimal control in quantum thermodynamics (including thermodynamic geometry~\cite{acconcia2015,zulkowski2015Quantum,scandi2019,deffner2020} and shortcuts to adiabaticity{~\cite{takahashi2017,guery2019,guery2022}}).

This review is organized as follows: we begin with examples of both experimental and theoretical model systems in section~\ref{Model Systems}, followed by a brief introduction to stochastic thermodynamics of heat, work, and entropy production in section~\ref{Thermodynamics}. Section~\ref{Full control} reviews recent progress exploiting optimal-transport theory to determine minimum-dissipation protocols under full control, yielding explicit solutions for the minimum-dissipation protocol under a strong-trap approximation in section~\ref{strong trap approximation} and allowing for constrained final control parameters in section~\ref{Constrained final control parameters}. Section~\ref{Parametric control} reviews parametric control, focusing on approximation methods in the fast (section~\ref{Fast control}), weak (section~\ref{Linear response}), and slow (section~\ref{Slow control}) limits. Applications to free-energy estimation are discussed in section~\ref{Free energy estimation} before comparing the performance of designed protocols in section~\ref{Comparison between control strategies} and finally concluding in section~\ref{Perspective and outlook} with a perspective and outlook for the study of optimal control in stochastic thermodynamics.

\section{Model systems}\label{Model Systems}

In this section we provide a brief introduction to a few paradigmatic model systems that motivate and guide the study of optimal control in stochastic thermodynamics. As discussed in section~\ref{Introduction}, the growth of stochastic thermodynamics coincides with the advent of new experimental techniques used to manipulate and measure single-molecule biophysical systems~\cite{Toyabe2011,toyabe2012,kawaguchi2014,svoboda1993,svoboda1994,kojima1997,hunt1994,greenberg2016,Laakso2008,norstrom2010,nagy2013}. First and foremost among these techniques are laser optical tweezers~\cite{bustamante2021,polimeno2018,moffitt2008} which can be used to trap microscopic Brownian systems.

The simplest experimental apparatus for studying stochastic thermodynamics is that of a microscopic bead trapped in an optical potential. From a theoretical perspective, this system is well approximated by continuous overdamped Brownian motion in a quadratic constraining potential. In these experiments, the center and stiffness of the trapping potential can be dynamically controlled to manipulate the system. With the use of feedback control, this experimental apparatus can be augmented to realize a virtual constraining potential of any form~\cite{kumar2018,kumar2019,gavrilov2017} and can, for example, be used to study fundamental bounds on information processing through bit erasure~\cite{jun2014,gavrilov2016}.

\begin{figure}
	\includegraphics[width=\linewidth]{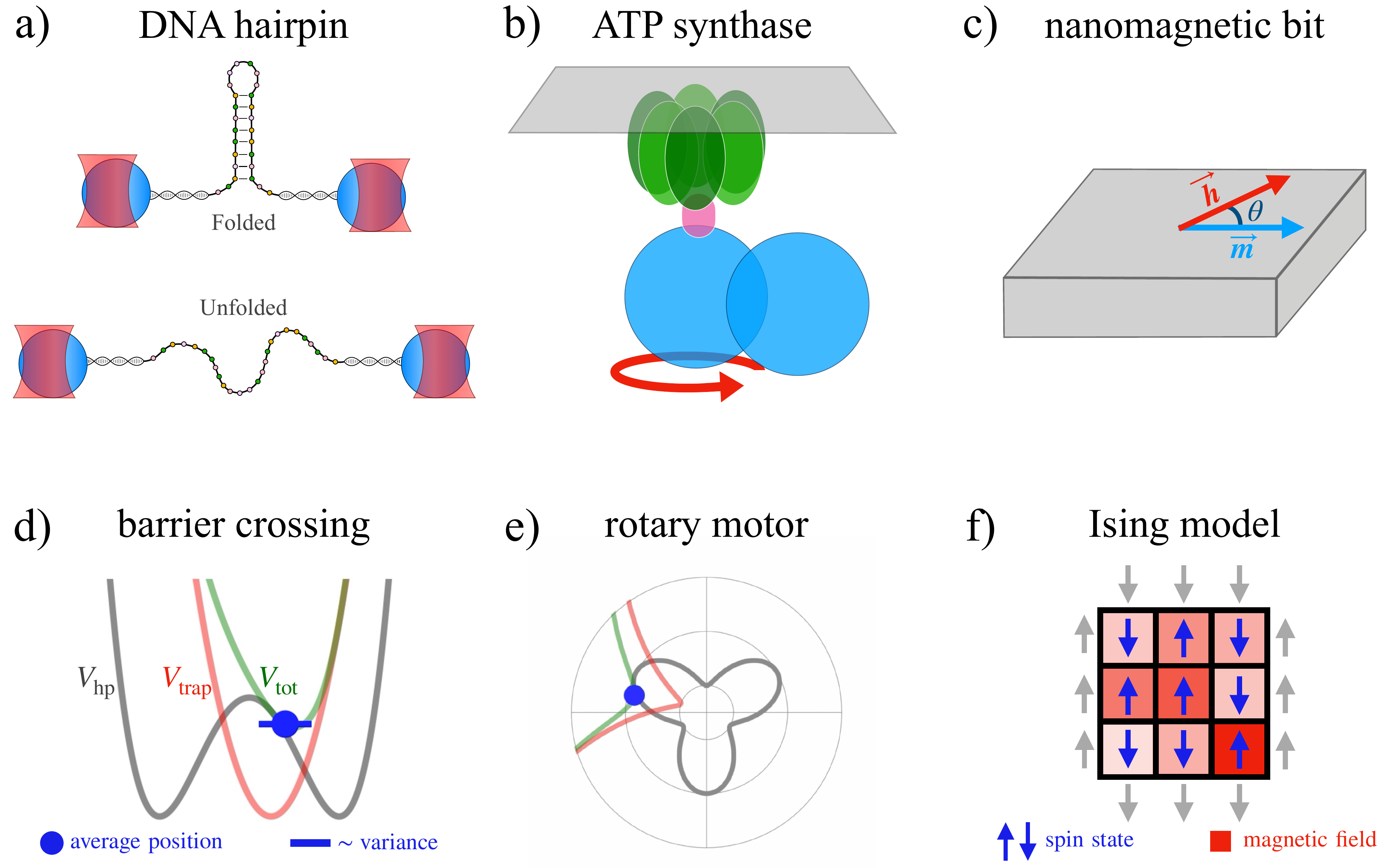}
	\caption{Model systems typical of stochastic thermodynamic control (top): a) DNA hairpin driven between folded and unfolded states by laser optical tweezers, b) ATP synthase driven by a magnetic trapping potential, c) nanomagnetic bit driven by an external magnetic field. Simplified theoretical descriptions (bottom) of the model systems in the top row: d) symmetric barrier-crossing model, e) Brownian rotary motor model, and f) nine-spin Ising model with independent magnetic fields applied to each spin.} 
	\label{fig_Model_Systems}
\end{figure}

Microscopic beads trapped by laser optical tweezers can be attached to biopolymers to probe their properties. For example, dual-trap optical tweezers can be used to fold and unfold DNA or RNA hairpins by modulating the separation between the trapping potentials (Fig.~\ref{fig_Model_Systems} a).\cite{liphardt2002,collin2005,bustamante2000,bustamante2003,woodside2006,neupane2017} Monitoring the position of the probe beads provides insight into the properties of the indirectly observed biopolymers. The simplest model representing this process is that of a driven barrier crossing~\cite{neupane2015}, where a Brownian system is dynamically driven over an energy barrier by a time-varying quadratic trapping potential (Fig.~\ref{fig_Model_Systems} d).\cite{Sivak2016,Blaber2022}

Magnetic traps can be used to probe the F1 component of the rotary motor ATP synthase (Fig.~\ref{fig_Model_Systems} b),\cite{Toyabe2011,toyabe2012,kawaguchi2014} which is driven periodically to synthesize ATP, an essential and portable energy source for the cell. Once again, microscopic beads are used to probe the properties of the molecular machine and can be dynamically driven (Fig.~\ref{fig_Model_Systems} e); however, the control differs from driven barrier crossing in that the driving is periodic. 

As a final example, consider a nanomagnetic bit characterized by its spin state or average magnetization (Fig.~\ref{fig_Model_Systems} c). By applying an external magnetic field, the system state can be driven from all spin-down to all spin-up, reversing the magnetization and resulting in a bit flip~\cite{hong2016}. This type of system is typically modeled with a discrete state space, e.g., the Ising model (Fig.~\ref{fig_Model_Systems} f), and optimal control of this system has been investigated~\cite{Rotskoff2015,Rotskoff2017,louwerse2022}. Due to the discrete state space, the properties of optimal control can differ from those for a system with a continuous state space.

Throughout this review we will use the model system of overdamped driven barrier crossing previously studied in Ref.~\onlinecite{Blaber2022} in order to give some intuition and examples for optimal control. The model consists of an overdamped Brownian particle in a double-well potential (symmetric for simplicity) constrained and driven by a quadratic trapping potential (Fig.~\ref{fig_Model_Systems} d). This system represents a simplified model of hairpin pulling experiments and Landauer erasure~\cite{Blaber2022,Proesmans2020,proesmans2020optimal}. The two-state nature of the system is also representative of activated processes such as chemical reactions, and the barrier crossing mimics the main features of experiments performed on ATP synthase, whose dynamics can be approximated as a series of barrier crossings~\cite{kawaguchi2014,Lucero2019,gupta2022}. This model is also typical of steered molecular-dynamics simulations, which use a time-dependent quadratic potential to drive reactions~\cite{Park2003,Park2004,Dellago2014}.

The total potential $V_{\rm tot}[x,x^{\rm c}(t),k(t)] = V_{\rm land}[x] + V_{\rm trap}[x,x^{\rm c}(t),k(t)]$ is the sum of the static hairpin potential $V_{\rm land}[x]$ and time-dependent trap potential $V_{\rm trap}[x,x^{\rm c}(t),k(t)]$ (shown schematically in Fig.~\ref{fig_Model_Systems}d). The hairpin potential is modeled as a static double well (symmetric for simplicity) with the two minima at $x = 0$ and $x = \Delta x_{\rm m}$ representing the folded and unfolded states~\cite{neupane2015,neupane2017,woodside2006,Sivak2016},
\begin{align}
V_{\rm land}(x) = E_{\rm B} \left[\left(\frac{2x-\Delta x_{\rm m}}{\Delta x_{\rm m}}\right)^2-1\right]^{2} \ ,
\label{double well}
\end{align}
for barrier height $E_{\rm B}$, distance $x_{\rm m}$ from the minimum to barrier, and distance $\Delta x_{\rm m}= 2x_{\rm m}$ between the minima. The system is driven by a quadratic trap
\begin{align}
V_{\rm trap}[x,x^{\rm c}(t),k(t)]= \frac{k(t)}{2}\left[x^{\rm c}(t)-x\right]^2 \ ,
\label{trap potential}
\end{align}
with time-dependent stiffness $k(t)$ and center $x^{\rm c}(t)$.

\section{Thermodynamics}\label{Thermodynamics}

In this review we focus on thermodynamics at the distribution level, which is generally described by dynamics of the form
\begin{align}
\frac{\partial p(\boldsymbol{r},t)}{\partial t}  = \mathcal{L}[\boldsymbol{r},t] \ p(\boldsymbol{r},t) \ ,
\label{General_dynamics}
\end{align}
governing the time evolution of a nonequilibrium probability distribution $p(\boldsymbol{r},t)$ over position vector $\boldsymbol{r}$ at time $t$ according to the time-evolution operator $\mathcal{L}[\boldsymbol{r},t]$. 

Continuous-space stochastic systems are described by the Fokker-Planck equation, which for overdamped dynamics has the time-evolution operator
\begin{align}
\mathcal{L}[\boldsymbol{r},t] = \nabla\cdot\boldsymbol{v}(\boldsymbol{r},t) + \boldsymbol{v}(\boldsymbol{r},t) \cdot\nabla \ ,
\label{Fokker Planck}
\end{align}
for mean local velocity~\cite{nakazato2021}
\begin{align}
\boldsymbol{v}(\boldsymbol{r},t) \equiv -D\nabla\left[\beta V_{\rm tot}(\boldsymbol{r},t) + \ln p(\boldsymbol{r},t)\right] \ ,
\label{velocity}
\end{align}
total potential $V_{\rm tot}$, diffusivity $D$, and $\beta \equiv (k_{\rm B}T)^{-1}$ for temperature $T$ and Boltzmann's constant $k_{\rm B}$. For a discrete-state system, $\mathcal{L}[\boldsymbol{r},t]$ is the transition rate matrix. For the example of driven barrier crossing, the total potential includes both the hairpin and trapping potential, and the time dependence arises from dynamic changes in the trap center and stiffness which drive the system over the barrier.

The average system energy is 
\begin{align}
    U = \int\md \boldsymbol{r} ~V(\boldsymbol{r},t)p(\boldsymbol{r},t) \ ,
\end{align}
and the rate of change in energy is 
\begin{align}
    \frac{\md U}{\md t} = \int\md \boldsymbol{r}\left[ \frac{\partial V(\boldsymbol{r},t)}{\partial t}p(\boldsymbol{r},t) + \frac{\partial p(\boldsymbol{r},t)}{\partial t}V(\boldsymbol{r},t)\right] \label{dU/dt}\ .
\end{align}
The first term quantifies work $W$ done on the system by an external agent controlling the potential $V(\boldsymbol{r},t)$ (e.g., an experimentalist dynamically driving a trapping potential) and the second quantifies heat flow $Q$ into the system from changes in the system distribution $p(\boldsymbol{r},t)$ (e.g., the system responding and relaxing towards a new equilibrium distribution in response to movement of the center and stiffness of the trapping potential):
\begin{align}
    \dot{W} &\equiv \int\md \boldsymbol{r} ~\frac{\partial V(\boldsymbol{r},t)}{\partial t}p(\boldsymbol{r},t) \ , \label{Work def}\\
    \dot{Q} &\equiv \int\md \boldsymbol{r}~\frac{\partial p(\boldsymbol{r},t)}{\partial t}V(\boldsymbol{r},t) \label{heat def}\ .
\end{align}
Throughout, a dot above a variable denotes the rate of change with respect to time. Substituting \eqref{Work def} and \eqref{heat def} into \eqref{dU/dt} gives the first law of thermodynamics: any change in system energy equals work and heat flows into the system.

Optimal control in thermodynamics is often discussed in terms of minimizing either entropy production or excess work incurred during the protocol. The entropy production as defined in this review will be used for periodic systems (e.g., ATP synthase), and when we have full control over the distribution (section~\ref{Full control}). The excess work is used for parametric control (section~\ref{Parametric control}) and model systems like DNA hairpins and nanomagnetic bits which are driven between control-parameter endpoints rather than periodically. 

To understand these two concepts, consider the nonequilibrium free energy 
\begin{subequations}
\begin{align}
    F_{\rm neq} &\equiv U - \beta^{-1}S \\ 
    &= \int\md \boldsymbol{r} ~\left[V(\boldsymbol{r},t)\, p(\boldsymbol{r},t) + \beta^{-1} \, p(\boldsymbol{r},t)\ln p(\boldsymbol{r},t)\right] \ , \label{Free Energy Def}
\end{align}
\end{subequations}
for dimensionless entropy
\begin{align}
    S \equiv -\int\md \boldsymbol{r}  ~p(\boldsymbol{r},t)\ln p(\boldsymbol{r},t) \ .
\end{align}
If the probabilities in~\eqref{Free Energy Def} are equilibrium distributions, then the nonequilibrium free energy reduces to the equilibrium free energy $F_{\rm eq}$. For an isothermal process the rate of change in free energy is
\begin{subequations}
\begin{align}
    \frac{\md F_{\rm neq}}{\md t} &= \frac{\md U}{\md t} - \beta^{-1}\frac{\md S}{\md t} \ , \\
    &= \dot{W} + \dot{Q} - \beta^{-1}\frac{\md S}{\md t} \ , \\
    &= \dot{W} -  \beta^{-1} \dot{S}_{\rm prod} \ .
\end{align}
\end{subequations}
The dimensionless entropy production is
\begin{subequations}
\begin{align}
\label{entropy_production_definition}
     \dot{S}_{\rm prod} &\equiv \frac{\md S}{\md t} + \frac{\md S_{\rm env}}{\md t} \\
     &= \frac{\md S}{\md t} - \beta \dot{Q} \\
     & \geq 0 \ , \label{Second Law}
\end{align}
\end{subequations}
for environmental entropy $S_{\rm env}$. Equation~\eqref{Second Law} follows from the second law of thermodynamics. Importantly, this definition of entropy production only accounts for dissipation during the protocol, so if the system is not in equilibrium at the end of the protocol, additional energy may be dissipated into the environment during subsequent relaxation.

A measure of energy dissipation which accounts for relaxation to equilibrium even after the protocol terminates is the excess work $W_{\rm ex} \equiv W - \Delta F_{\rm eq}$, the amount of work done in excess of the equilibrium free-energy difference. The excess work and entropy production are related by
\begin{align}
\label{excess_work_and_entropy_production}
    W_{\rm ex} = \Delta F_{\rm neq}-\Delta F_{\rm eq} + \beta^{-1}\Delta S_{\rm prod} \ ,
\end{align}
for net entropy production $\Delta S_{\rm prod} \equiv \int_{0}^{\Delta t} \md t \, \dot{S}_{\rm prod}$, nonequilibrium free-energy difference $\Delta F_{\rm neq}$ between the initial $p(\boldsymbol{r},0)$ and final $p(\boldsymbol{r},\Delta t)$ distributions, and equilibrium free-energy difference $\Delta F_{\rm eq}$ between the initial and final equilibrium distributions. 

Equation~\eqref{excess_work_and_entropy_production} clarifies the distinction between entropy production and excess work: if both the initial and final states of the system are at equilibrium, the two quantities are equal, otherwise the difference between the two equals the difference between the nonequilibrium and equilibrium free-energy changes. If the system is allowed to relax to equilibrium, then this excess energy is dissipated into the environment, resulting in additional entropy production not accounted for in the present definition~\eqref{entropy_production_definition}, so that the total entropy production for such a process is~\eqref{entropy_production_definition} plus the entropy production from the subsequent relaxation. In contrast, the excess work always includes the energy dissipated into the environment from the system relaxing towards equilibrium after the protocol terminates. Essentially, quantifying dissipation by the entropy production in~\eqref{entropy_production_definition} assumes one can harness the nonequilibrium free energy at the conclusion of the protocol to perform a useful task (generally true for periodically driven systems like ATP synthase), while excess work quantifies dissipation when all the excess free energy is dissipated into the environment after the protocol terminates (generally true for two-state barrier crossings like hairpin experiments).

\section{Full control}\label{Full control}

For continuous-state systems, complete control over the shape of the potential $V(\boldsymbol{r},t)$ grants full control over the probability distribution $p(\boldsymbol{r},t)$, which can considerably simplify the optimization process and allows us to exploit known results from optimal-transport theory~\cite{Aurell2011,nakazato2021,ito2022}. Optimal transport describes the most efficient methods to move mass (e.g., a pile of sand) from one location to another; this is useful for describing methods that minimize dissipation in transporting probability from an initial to a final distribution~\cite{villani2009}.

Since the final distribution is constrained, the energy dissipated into the environment throughout the protocol is determined by the average entropy produced in driving from initial probability distribution $p(\boldsymbol{r},0)$ to final probability distribution $p(\boldsymbol{r},\Delta t)$ as~\cite{nakazato2021}
\begin{align}
\label{OT_entropy}
\Delta S_{\rm prod} = \frac{1}{D} \int_{0}^{\Delta t} \md t \ \langle \boldsymbol{v}(\boldsymbol{r},t)\cdot \boldsymbol{v}(\boldsymbol{r},t)\rangle \ .
\end{align}
Angle brackets $\langle\cdots \rangle$ denote an average over $p(\boldsymbol{r},t)$.

Expressing the entropy production in this form makes precise the connections between optimal-transport theory and minimum-dissipation protocols. In optimal-transport theory a common measure of the distance between two distributions is the $L_{2}$-Wasserstein distance, defined in the Benamou and Brenier dual representation as~\cite{benamou2000}
\begin{align}
    \mathcal{W}(p_{0}, p_{\Delta t})^2 \equiv \min_v \int_{0}^{\Delta t} \md t \ \langle \boldsymbol{v}(\boldsymbol{r},t)\cdot \boldsymbol{v}(\boldsymbol{r},t)\rangle \ .
\end{align}
Therefore, the entropy production is bounded by the squared $L_{2}$-Wasserstein distance between initial ($p_0$) and final ($p_{\Delta t}$) probability distributions~\cite{nakazato2021}:
\begin{align}
\label{entropy_production_bound}
    \Delta S_{\rm prod} \geq \frac{\mathcal{W}\left(p_0, p_{\Delta t}\right)^2}{D \Delta t} \ .
\end{align}
This allows us to exploit existing procedures from optimal-transport theory to determine protocols that minimize entropy production~\cite{villani2009,santambrogio2015}. Notably, the exact solution is known in two situations: one-dimensional systems and Gaussian probability distributions (section~\ref{Exact solutions}).

Extending minimum-dissipation full control and the connections to optimal-transport theory to more general forms of dynamics (e.g., discrete state spaces) is a rapidly advancing area of active research.\cite{dechant2022,dechant2022geometric,yoshimura2022,zhong2022,van2022,van2022Topological}

\subsection{Exact solutions}\label{Exact solutions}
For a one-dimensional system $\boldsymbol{r} = x$, the entropy production~\eqref{OT_entropy} of the optimal-transport process can be simplified considerably as~\cite{Aurell2011,Abreu2011,Zhang2019,zhang2020,Proesmans2020,proesmans2020optimal}
\begin{align}
    \Delta S_{\rm prod} \geq \frac{1}{D\Delta t}\int_{0}^{1}\md y \left[\mathcal{Q}_{\rm f}(y)-\mathcal{Q}_{\rm i}(y)\right]^2 \ , 
\end{align}
where $\mathcal{Q}_{\rm f}$ and $\mathcal{Q}_{\rm i}$ are the final and initial quantile functions (inverse cumulative distribution functions)~\cite{Blaber2022}. The entropy production is minimized if the quantiles are linearly driven between their fixed initial and final values~\cite{Zhang2019,zhang2020,Proesmans2020,proesmans2020optimal}; from this the probability distribution can be computed, and then the Fokker-Planck equation inverted to determine the potential $V_{\rm tot}(x,t)$ to be applied to achieve the control that minimizes the entropy production:
\begin{align}
    V_{\rm tot}(x,t) = -\beta^{-1}\ln p(x,t) + (\beta D)^{-1}\int_{-\infty}^{x}\md x' ~\frac{\int_{-\infty}^{x'}\md x''~\frac{\partial p(x'',t)}{\partial t} }{p(x',t)} \ .
\end{align}
Although this calculation is often analytically intractable, it is straightforward to compute numerically for any probability distribution.

If the initial and final distributions are Gaussian, $p(\boldsymbol{r},t) = \mathcal{N}(\mu_t,\Sigma_t)$ for time-dependent mean $\mu_t$ and covariance $\Sigma_t$, the entropy-production bound~\eqref{entropy_production_bound} is~\cite{Olkin1982,dechant2019,abiuso2022}
\begin{align}
    \label{Gaussian Entropy}
	\Delta S_{\rm prod} \geq \frac{1}{D\Delta t}\bigg\{ \Delta\boldsymbol{\mu}^2 + {\rm Tr}\left[\Sigma_{0} + \Sigma_{\Delta t} - 2(\Sigma_{\Delta t}^{\frac{1}{2}}\Sigma_{0}\Sigma_{\Delta t}^{\frac{1}{2}})^{\frac{1}{2}}\right]\bigg\} \ ,
\end{align}
with subscripts $0$, $t$, and $\Delta t$ respectively denoting the initial, time-dependent, and final values of the corresponding variable. Equality is achieved and the entropy production is minimized when following the optimal-transport map between the initial and final distributions, which for Gaussian distributions is completely specified by the mean and covariance:
\begin{subequations}
\begin{align}
\label{optimal mean}
\boldsymbol{\mu}_{t} &= \boldsymbol{\mu}_{0}+\Delta \boldsymbol{\mu}\frac{t}{\Delta t} \\
\Sigma_{t} &= \left[\left(1-\frac{t}{\Delta t}\right)I +\frac{t}{\Delta t}C \right]\Sigma_{0}\left[\left(1-\frac{t}{\Delta t}\right)I+\frac{t}{\Delta t}C\right] \ .
\label{optimal variance}
\end{align}
\end{subequations}
Here $\Delta \boldsymbol{\mu} \equiv \boldsymbol{\mu}_{\Delta t} - \boldsymbol{\mu}_{0}$ is the total change in mean position, $I$ is the identity matrix, and $C \equiv \Sigma_{\Delta t}^{\frac{1}{2}}(\Sigma_{\Delta t}^{\frac{1}{2}}\Sigma_{0}\Sigma_{\Delta t}^{\frac{1}{2}})^{-\frac{1}{2}}\Sigma_{\Delta t}^{\frac{1}{2}}$ reduces in 1D to the ratio of final and initial standard deviations. If the covariance matrix $\Sigma$ is diagonal, then \eqref{optimal variance} implies
\begin{align}
    \Sigma_{t}^{\frac{1}{2}} &= \Sigma_{0}^{\frac{1}{2}}+\Delta \Sigma^{\frac{1}{2}}\frac{t}{\Delta t} \ ,
    \label{optimal variance diagonal}
\end{align}
with $\Delta \Sigma^{\frac{1}{2}} \equiv \Sigma^{\frac{1}{2}}_{\Delta t}-\Sigma^{\frac{1}{2}}_{0}$. Thus for diagonal covariance the optimal-transport process linearly drives the standard deviation between its endpoint values. For a detailed description of minimum-dissipation protocols for general multidimensional Gaussian distributions see Ref.~\onlinecite{abiuso2022}.

In both solvable cases (1D and Gaussian) the general design principle is that the minimum-dissipation protocol linearly drives the quantiles between specified initial and final values. Linearly driving the quantiles of the probability distribution can be used as a guiding principle for designing more general minimum-dissipation protocols and arises independently for parametric control~\cite{Blaber2022}.

\subsection{Strong-trap approximation}\label{strong trap approximation}

In this section we review how to exploit the known full-control solution for Gaussian distributions in order to design minimum-dissipation protocols for strong trapping potentials on any arbitrary underlying energy landscape, initially described in Ref.~\onlinecite{Blaber2022_strong}. 

We assume that the total potential $V_{\rm tot}[\boldsymbol{r},\boldsymbol{r}_{\rm c}(t),K(t)] = V_{\rm land}(\boldsymbol{r}) + V_{\rm trap}[\boldsymbol{r},\boldsymbol{r}_{\rm c}(t),K(t)]$ can be separated into a time-independent component $V_{\rm land}(\boldsymbol{r})$ (the underlying energy landscape) and a quadratic trapping potential
\begin{align}
V_{\rm trap}[\boldsymbol{r},\boldsymbol{r}_{\rm c}(t),K(t)]=\frac{1}{2}\left[\boldsymbol{r}-\boldsymbol{r}_{\rm c}(t)\right]^{\top} K(t)\left[\boldsymbol{r}-\boldsymbol{r}_{\rm c}(t)\right] \ .
\end{align}
The position of the system is denoted by the vector $\boldsymbol{r}$, $K$ is the symmetric stiffness matrix, and superscript $\top$ is the vector transpose. For a strong trapping potential, the time-independent component can be expanded up to second order about the mean particle position $\boldsymbol{\mu}$:
\begin{align}
V_{\rm land}(\boldsymbol{r}) \approx V_{\rm land}(\boldsymbol{\mu}) + (\boldsymbol{r}-\boldsymbol{\mu})^{\top}\nabla V_{\rm land}(\boldsymbol{\mu}) +\frac{1}{2}(\boldsymbol{r}-\boldsymbol{\mu})^{\top} \nabla\nabla^{\top} V_{\rm land}(\boldsymbol{\mu}) (\boldsymbol{r}-\boldsymbol{\mu}) \ .
\end{align}
Under these assumptions, the probability distribution can be approximated as Gaussian, $p(\boldsymbol{r},t) \approx \mathcal{N}(\boldsymbol{\mu}_{t},\Sigma_{t})$. Since the distribution is Gaussian we can use the results described by~\eqref{optimal mean} and~\eqref{optimal variance}. 

To achieve the mean and covariance of~\eqref{optimal mean} and~\eqref{optimal variance}, the trap center and stiffness must respectively satisfy
\begin{subequations}
\label{eq:Optima}
\begin{align}
\boldsymbol{\lambda}_t =& \boldsymbol{\mu}_{t} + K_{t}^{-1}\left[\frac{\Delta \boldsymbol{\mu}}{\beta D \Delta t}+\nabla V_{\rm land}(\boldsymbol{\mu}_t)\right] \ , \label{optimal center} \\
K_{t}=& \frac{1}{\beta}\Sigma_{t}^{-1} - \frac{1}{\beta D}\int_{0}^{\infty}\md\nu \  e^{-\nu\Sigma_t}\frac{\md \Sigma_t}{\md t}e^{-\nu\Sigma_t} - \nabla\nabla^{\top} V_{\rm land}(\boldsymbol{\mu}_t) \ . \label{optimal stiffness general}
\end{align}
\end{subequations}
If $\Sigma$ is diagonal, then $\Sigma_t$ is given by \eqref{optimal variance diagonal}, the integral in~\eqref{optimal stiffness general} can be evaluated, and the trap stiffness obeys
\begin{align}
    K_{t} &=\nabla\nabla^{\top} V_{\rm land}(\boldsymbol{\mu}_t) + \left(\frac{1}{\beta}I- \frac{\Delta \Sigma^{\frac{1}{2}} }{2\beta D\Delta t}\right)\Sigma_{t}^{-1} . 
\end{align}

Equations~\eqref{optimal center} and~\eqref{optimal stiffness general} are explicit equations that minimize dissipation at any driving speed provided the trap is sufficiently strong. We emphasize that explicit solutions for the minimum-dissipation protocol are rarities; typically solutions require numerically solving differential equations or inverting the Fokker-Planck equation~\cite{Aurell2011,Proesmans2020,proesmans2020optimal}.

For our example system of driven barrier crossing with equal initial and final covariance, the protocol designed by \eqref{optimal center} slows down movement of the trap center as it crosses the barrier to compensate for the force due to the hairpin potential, and \eqref{optimal stiffness} tightens the trap as it crosses the barrier in order to counteract the negative curvature of the hairpin potential. Slowing down and tightening the trapping potential as the system crosses a barrier appears to be a general feature and results independently from other approximation methods~\cite{Blaber2022,Blaber2022_strong}.

To achieve periodic driving we constrain the final covariance matrix after one period (during which the mean completes one rotation) to equal the initial. To minimize dissipation, the standard deviation is linearly changed between the endpoints~\eqref{optimal variance}, implying for a periodic system that the standard deviation (and hence covariance) remains unchanged throughout the protocol. This is achieved when the effective stiffness is constant, i.e.,
\begin{align}
    K_{t}= K_{0}+\nabla \nabla^{\top} V_{\rm land}(\boldsymbol{\mu}_{0}) -\nabla \nabla^{\top} V_{\rm land}(\boldsymbol{\mu}_{t}) \ .
\label{optimal stiffness}
\end{align}
If in each rotation the mean travels a distance $\Delta\boldsymbol{\mu}_{\rm rot}$ in time $\Delta t_{\rm rot}$, the resultant entropy production is
\begin{align}
\Delta S_{\rm prod} =  \frac{\Delta\boldsymbol{\mu}_{\rm rot}^{2} }{D\Delta t_{\rm rot}} \ ,
\label{entropy production at constant variance}
\end{align}
that of an overdamped system moving at constant velocity against viscous Stokes drag; i.e., the minimum-dissipation protocol has perfect \emph{Stokes efficiency}~\cite{wang2002}.

\subsection{Constrained final control parameters}\label{Constrained final control parameters}

The techniques described in section~\ref{Full control} minimize the entropy production for constrained final distributions. Constraining the final distribution allows us to describe periodically driven systems like models inspired by ATP synthase; however, often the end state is instead constrained by the final control-parameter values. For example, in hairpin experiments the end state is typically defined by the trap separation, and the system is typically allowed to equilibrate between subsequent unfolding/folding protocols. Since the system is allowed to equilibrate between protocols, any excess energy remaining at the end of the protocol is dissipated as heat into the environment, and the entropy production as defined by \eqref{entropy_production_definition} does not equal the total dissipation.

In this case the relevant thermodynamic quantity is the work~\eqref{excess_work_and_entropy_production},
\begin{align}
    W = \Delta F_{\rm neq} + \Delta S_{\rm prod} \ ,
\end{align}
which accounts for the excess free energy dissipated into the environment after the protocol terminates. Work can be minimized by optimizing over the final distribution subject to constrained final control-parameter values~\cite{Blaber2022_strong,Blaber2022}. In general this minimization must be performed numerically, which is straightforward for one-dimensional systems, but becomes computationally intensive for higher-dimensional systems.

For the special case of Gaussian distributions, the optimization procedure simplifies considerably since optimizing over a distribution reduces to optimizing over means and covariances. The average work for the minimum-dissipation protocol in the strong-trap approximation is~\cite{Blaber2022_strong}
\begin{align}
    W  = & \frac{1}{2}{\rm Tr}\left\{K\left[\Sigma+(\boldsymbol{\mu}-\boldsymbol{\lambda})(\boldsymbol{\mu}-\boldsymbol{\lambda})^{\top}\right]\right\}_{0}^{\Delta t}+V_{\rm land}(\boldsymbol{\mu})\big|_0^{\Delta t} + \frac{1}{2}{\rm Tr}\left[\nabla\nabla^{\top} V_{\rm land}(\boldsymbol{\mu})\Sigma\right]_{0}^{\Delta t} + \beta^{-1}\Delta S_{\rm prod}^{\rm min} \ ,
    \label{Optimal Work}
\end{align}
with ${\rm Tr}$ the trace and $\Delta S_{\rm prod}^{\rm min}$ the lower bound in \eqref{Gaussian Entropy}. 

To find the protocol that minimizes work for constrained final control parameters, we minimize \eqref{Optimal Work} with respect to the final mean $\mu_{\Delta t}$ and covariance $\Sigma_{\Delta t}$, for fixed final trap center $\boldsymbol{\lambda}_{\Delta t}$ and stiffness $K_{\Delta t}$. For a flat energy landscape (for which the strong-trap approximation is exact) the optimal mean and covariance can be solved analytically; e.g., for equal initial and final covariance, the final mean is
\begin{align}
\boldsymbol{\mu}_{\Delta t} = \boldsymbol{\mu}_0 + \left(\frac{2K^{-1}}{\beta D\Delta t} + I\right)^{-1}[\boldsymbol{\lambda}_{\Delta t} - \boldsymbol{\mu}_0] \ .
\label{Optimal Final Mean}
\end{align}
In some more general cases (e.g., energy landscapes represented by low-order polynomials), \eqref{Optimal Work} can also be minimized analytically, and in general can be solved numerically with relative ease.

\section{Parametric control}\label{Parametric control}

While full-control solutions are convenient when possible, many applications do not permit sufficient control to fully constrain the probability distribution throughout the entire protocol. In such cases the controller is constrained by a finite set of control parameters $\boldsymbol{\lambda}(t)$ which can be used to drive the system. Since there are insufficient control parameters to fully control the probability distribution, the endpoints of the protocol are constrained by the control-parameter values rather than the distribution. Therefore, we focus on optimizing the excess work, which is an accurate measure of dissipation provided the system equilibrates after the protocol terminates.

In this section we assume the control is the result of time-dependent control parameters $\boldsymbol{\lambda}(t)$, in which case the excess work~\eqref{Work def} is
\begin{align}
\label{parametric work}
    \langle W_{\rm ex} \rangle_{\Lambda} ~= -\int_{0}^{\Delta t} {\rm d} t~  \langle \delta {\bf f}^{\top}(t) \rangle_{\Lambda} \dot{\boldsymbol{\lambda}}(t)\ ,
\end{align}
where we have defined the conjugate force ${\bf f} \equiv - \partial V_{\rm tot}/\partial \boldsymbol{\lambda}$ to express the work explicitly in terms of the control parameters, and $\delta$ denotes a difference from the equilibrium average. In this section we denote a nonequilibrium average (dependent on the entire protocol history) as $\langle \cdot \rangle_{\Lambda}$ and an equilibrium average (at the current control parameter value $\boldsymbol{\lambda}(t)$) as $\langle \cdot \rangle_{\boldsymbol{\lambda}(t)}$. Since the control-parameter protocol is externally specified, the only unknown is $\langle \delta {\bf f}^{\top}(t) \rangle_{\Lambda}$. This quantity is particularly difficult to deal with since the nonequilibrium average depends on the entire protocol history. 

For example, Ref.~\onlinecite{Schmiedl2007} showed that---even in one dimension---optimization requires solving the nonlocal Euler-Lagrange equation. Therefore, there are very few cases where the minimum-work protocol can be determined analytically. Promising approaches for obtaining general solutions include optimal-transport theory with limited control~\cite{zhong2022} and advanced numerical techniques~\cite{Then2008,gingrich2016,engel2022}.

The minimum-work protocol can be determined exactly for a Brownian particle in a harmonic trap, in which case the optimal protocol, originally derived in one dimension~\cite{Schmiedl2007}, is identical to the full-control solution for Gaussian distributions described in section~\ref{Constrained final control parameters}. This exact solution serves as a window into the properties of minimum-dissipation protocols and gives us considerable insight into what to expect from optimized protocols: e.g., the minimum-dissipation protocols have control-parameter jumps at the start and end but remain continuous in between.

Although exact solutions are nice when possible, since general solutions are intractable we turn to approximate methods to gain insight into the general properties of minimum-dissipation protocols. The first approximation has already been described in section~\ref{strong trap approximation} and is valid for strong trapping potentials applied to overdamped systems~\cite{Blaber2022_strong}. In the following sections, we fill in the remaining limits of fast, weak, and slow control in order to describe optimal control for any system with any given control parameters and design interpolated protocols.

\subsection{Fast control}\label{Fast control}

Until recently, although it was generally suspected, it was not definitively known whether the jumps in the minimum-dissipation protocols are a general feature. By taking the fast-driving limit, it has been shown that the minimum-dissipation protocol is a step function, jumping from and to specified initial and final control-parameter endpoints to spend the entire intervening protocol duration at the control-parameter values that maximize the short-time power savings.\cite{Blaber2021} We refer to the minimum-dissipation protocols in the fast limit as short-time efficient protocols, or STEP.

In the fast limit, the excess work approaches that of an instantaneous protocol, which spends no time relaxing towards equilibrium and requires excess work proportional (up to a factor of $\beta$) to the relative entropy $D(\pi_{\rm i}||\pi_{\rm f})$ between the respective initial and final equilibrium distributions, $\pi_{\rm i}$ and $\pi_{\rm f}$~\cite{Blaber2021}. Spending a short duration $\Delta t$ relaxing towards equilibrium throughout the protocol results in saved work $W_{\rm save} \equiv \beta^{-1} D(\pi_{\rm i}||\pi_{\rm f}) -W_{\rm ex}$ compared to an instantaneous protocol, which can be approximated as~\cite{Blaber2021}
\begin{align}
\langle W_{\rm save}\rangle_{\Lambda} \approx  \int_{0}^{\Delta t}\md t  \, {\bf R}_{\boldsymbol{\lambda}_{\rm i}}^{\top}[\boldsymbol{\lambda}(t)] \, 
[\boldsymbol{\lambda}_{{\rm f}} - \boldsymbol{\lambda}(t)] \ ,
\label{Excess work approx}
\end{align}
in terms of the initial force-relaxation rate
\begin{align}
	{\bf R}_{\boldsymbol{\lambda}_{\rm i}}[\boldsymbol{\lambda}(t)] &\equiv \frac{\md\langle {\bf f} \rangle_{\boldsymbol{\lambda}_{\rm i}}}{\md t}\bigg|_{\boldsymbol{\lambda}(t)} \ ,
	\label{Rate Function}
\end{align}
the rate of change of the initial mean conjugate forces at the current control-parameter values. 

The saved work is maximized by the short-time efficient protocol (STEP) which spends the entire duration at the intermediate control-parameter value that maximizes the short-time power savings
\begin{align}
P_{\rm save}^{\rm st}(\boldsymbol{\lambda}) \equiv {\bf R}_{\boldsymbol{\lambda}_{\rm i}}^{\top}(\boldsymbol{\lambda})(\boldsymbol{\lambda}_{{\rm f}} - \boldsymbol{\lambda}) \ ,
\label{eq:power_savings}
\end{align}
satisfying
\begin{align}
&\frac{\partial P_{\rm save}^{\rm st}(\boldsymbol{\lambda})}{\partial \boldsymbol{\lambda}}\bigg|_{\boldsymbol{\lambda}^{\rm STEP}}  = 0 \ .
\label{STEP}
\end{align}
The STEP achieves this by two instantaneous control-parameter jumps: one at the start from the initial value to the optimal value $\boldsymbol{\lambda}^{\rm STEP}$, and one at the end from $\boldsymbol{\lambda}^{\rm STEP}$ to the final value.

The optimal protocol described in this section is general, only assuming the protocol is fast compared to the system's natural relaxation time. Additionally, for quadratic time-dependent potentials (such as the driven barrier-crossing model system) or affine control ($V_{\rm tot}(\boldsymbol{r}, \lambda)=V_0(\boldsymbol{r})+\lambda V_1(\boldsymbol{r})$ where $\lambda$ linearly modulates the strength of an auxiliary potential $V_1(\boldsymbol{r})$ added to the base potential $V_0(\boldsymbol{r})$) the STEP value is always halfway between the control-parameter endpoints~\cite{Blaber2022,zhong2022}. Finally, since the initial force-relaxation rate~\eqref{Rate Function} is an equilibrium average and the STEP value is given by a point in control-parameter space, the STEP value is simple to determine. The STEP and the strong-trap optimal protocol are the simplest protocols to determine and can be calculated analytically in many cases or numerically evaluated with relative ease.

\subsection{Linear response}\label{Linear response}

Linear-response theory can be used to determine the minimum-dissipation protocol for weak perturbations and performs relatively well at any driving speed, well beyond its strict range of validity~\cite{kamizaki2022,bonanca2018}. For fast driving, the minimum-dissipation protocols determined from linear-response theory have jumps at the start and end of the protocol. 

As mentioned in section~\ref{Parametric control}, the central quantity that needs to be determined to perform any optimization of the excess work is the nonequilibrium average force. In linear response, the deviation of the nonequilibrium average force from equilibrium is approximated by the integrated equilibrium force covariance 
\begin{align}
\label{LR average force}
    \langle \delta f_{j}(t) \rangle_{\Lambda} \approx -\int_0^{t} {\rm d} t' ~ \langle \delta f_{j}(t-t') ~ \delta f_{\ell}(0)\rangle_{\lambda(t)}~\dot{\lambda}_{\ell}(t') \ .
\end{align}
This greatly simplifies the optimization procedure, since the equilibrium average depends only on the current control-parameter value and not on the entire history of control. Substituting this into the excess work~\eqref{parametric work} gives
\begin{align}
    \langle W_{\rm ex} \rangle_{\Lambda} ~\approx ~\int_{0}^{\Delta t} {\rm d} t \int_0^{t} {\rm d} t' ~ ~ \dot{\lambda}_{j}(t)~  \langle \delta f_{j}(t-t') ~ \delta f_{\ell}(0)\rangle_{\lambda(t)}~\dot{\lambda}_{\ell}(t') \ ,
\end{align}
which can be optimized directly by numerical methods and can perform well at any driving speed~\cite{bonanca2018,kamizaki2022}.

\subsection{Slow control}\label{Slow control}

The next approximation we consider is valid for slow near-equilibrium processes. This approach generalizes the paradigm of thermodynamic geometry to stochastic thermodynamics~\cite{Crooks2007,OptimalPaths}. The generalized friction tensor endows the space of thermodynamic states with a Riemannian metric where minimum-dissipation protocols correspond to geodesics of the friction tensor. This method is widely applicable, yields a relatively simple prescription for determining minimum-dissipation protocols, and has been extended to more general settings and different forms of control. The protocols determined from this method are continuous; however, we know from exact solutions that minimum-dissipation protocols can have jump discontinuities~\cite{Schmiedl2007}, which are never optimal within the geometric framework of slow control. This arises from the slow near-equilibrium approximation used; indeed, in the limit of a slow protocol the jumps in the exact optimal protocol become negligible.

In addition to the linear-response approximation, we assume the control parameters are driven slowly compared to the system's natural relaxation timescale, so the approximation for the nonequilibrium average force~\eqref{LR average force} simplifies to
\begin{align}
    \langle \delta f_{j}(t) \rangle_{\Lambda} \approx -\dot{\lambda}_{\ell}(t)\int_0^{\infty} {\rm d} t' ~ \langle \delta f_{j}(t') ~ \delta f_{\ell}(0)\rangle_{\lambda(t)} \ .
\end{align}
Substituting into~\eqref{parametric work} yields the leading-order contribution to the excess work~\cite{OptimalPaths}:
\begin{align}
\label{LR excess work}
\langle W_{\rm ex}\rangle_{\Lambda} \approx \int_{0}^{\Delta t}\md t \ \frac{\md {\boldsymbol{\lambda}}^{\top}}{\md t} \ \zeta[\boldsymbol{\lambda}(t)] \ \frac{\md {\boldsymbol{\lambda}}}{\md t} \ ,
\end{align}
in terms of the generalized friction tensor with elements
\begin{align}
\zeta_{j \ell}(\lambda) \equiv \beta \int_0^{\infty} \md t \, \langle \delta f_{j}(t) \delta f_{\ell}(0)\rangle_{\boldsymbol{\lambda}} \ . 
\label{friction}
\end{align}
In analogy with fluid dynamics, this rank-two tensor is the \emph{Stokes' friction}, since it produces a drag force that depends linearly on velocity.

$\zeta_{j \ell}$ is the Hadamard product $\beta \langle\delta f_{j} \delta f_{\ell}\rangle_{\boldsymbol{\lambda}} \circ \tau_{j \ell}$ of the conjugate-force covariance (the force fluctuations) and the integral relaxation time
\begin{align}
\label{relax1}
\tau_{j \ell} \equiv \int_0^{\infty} \md t \, \frac{\langle \delta f_{j}(t) \delta f_{\ell}(0)\rangle_{\boldsymbol{\lambda}}}{\langle \delta f_{j} \delta f_{\ell}\rangle_{\boldsymbol{\lambda}}} \ ,
\end{align}
the characteristic time for these fluctuations to die out.

For overdamped dynamics, the friction can be calculated directly from the total energy as~\cite{zulkowski2015}
\begin{align}
\zeta_{j\ell}(\boldsymbol{\lambda}) = \int_{-\infty}^{\infty} \md x 
\, 
\frac{\partial_{\lambda_{j}}\Pi_{\rm eq}(x,\boldsymbol{\lambda})\partial_{\lambda_{\ell}}\Pi_{\rm eq}(x,\boldsymbol{\lambda})}{\pi_{\rm eq}(x,\boldsymbol{\lambda})}
\ ,
\label{CDF friction}
\end{align}
where $\Pi_{\rm eq}(x,\boldsymbol{\lambda}) \equiv \int_{-\infty}^{x}\md x'\pi_{\rm eq}(x',\boldsymbol{\lambda})$ is the equilibrium cumulative distribution function, $\partial_{\lambda_{j}}$ is the partial derivative with respect to $\lambda_{j}$, and $\pi_{\rm eq}(x',\boldsymbol{\lambda}) = \exp[-\beta V_{\rm tot}(x,\boldsymbol{\lambda})]~/~\int\md x \exp[-\beta V_{\rm tot}(x,\boldsymbol{\lambda})]$ is the equilibrium probability distribution.

Within the slow-protocol approximation, the excess work is minimized by a protocol with constant excess power~\cite{OptimalPaths}. For a single control parameter, this amounts to proceeding with velocity $\md \lambda^{\rm LR}/\md t \propto \zeta(\lambda)^{-1/2}$, which when normalized to complete the protocol in a fixed allotted time $\Delta t$, gives 
\begin{align}
\label{lambdaoptdot}
\frac{\md \lambda^{\rm LR} }{\md t}= \frac{\Delta \lambda}{\Delta t}\frac{\overline{{\zeta}^{1/2}}}{\sqrt{\zeta(\lambda)}} \ ,
\end{align}
where the overline denotes the spatial average over the naive (linear) path between the control-parameter endpoints.

For multidimensional control, the minimum-dissipation protocol solves the Euler-Lagrange equation
\begin{equation}
\zeta_{j \ell}\frac{\md^2\lambda_{\ell}}{\md t^2}+ \frac{\partial\zeta_{j \ell}}{\partial \lambda_{m}} \frac{\md\lambda_{\ell}}{\md t}\frac{\md\lambda_{m}}{\md t} = \frac{1}{2}\frac{\partial\zeta_{\ell m}}{\partial \lambda_{j}} \frac{\md\lambda_{\ell}}{\md t}\frac{\md\lambda_{m}}{\md t} \ ,
\label{Euler-Lagrange}
\end{equation}
where we have adopted the Einstein convention of implied summation over all repeated indices. 

\begin{figure}
	\includegraphics[width=\linewidth]{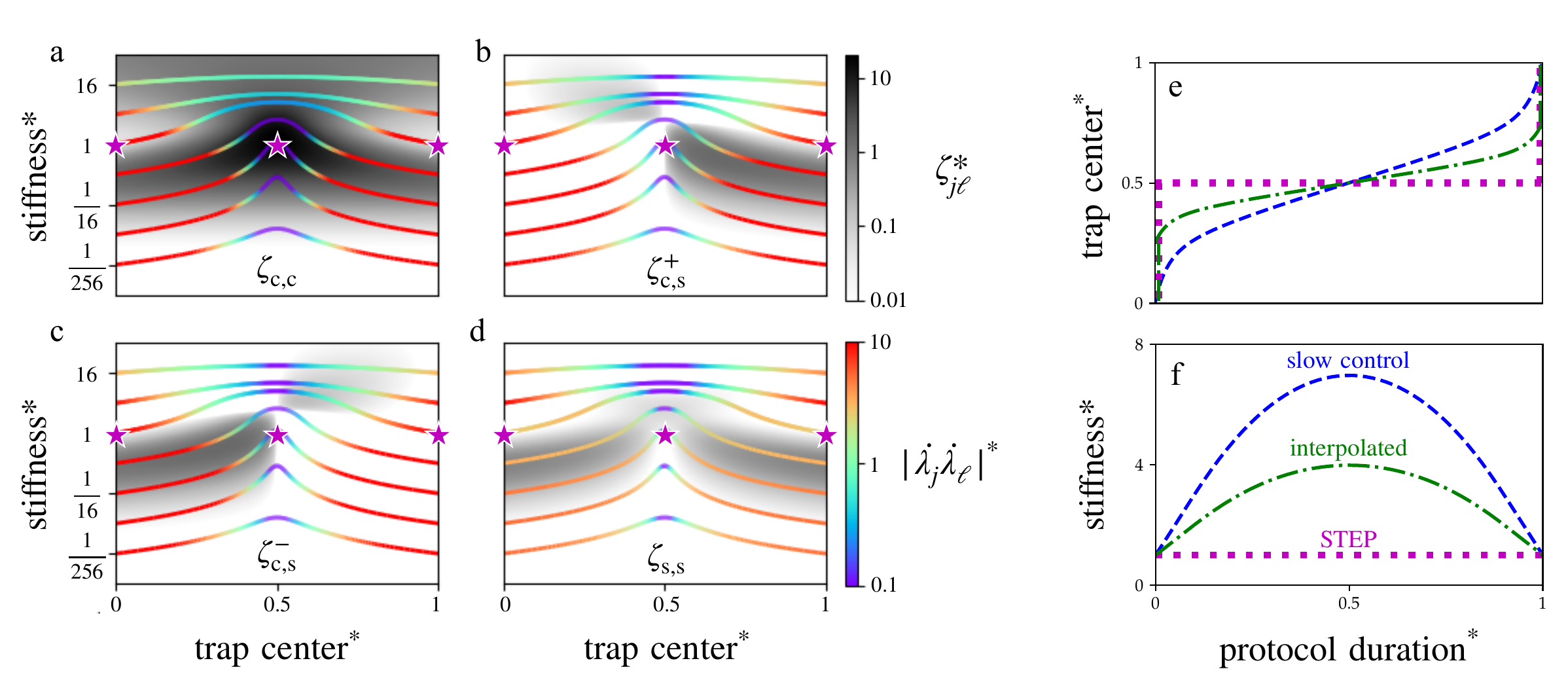}
	\caption{Geodesics and components of the friction matrix used to design two-dimensional linear-response protocols. Grayscale heatmap: components of the friction as a function of the (dimensionless) trap center$^*$ $x^{\rm c}/{\Delta x_{\rm m}}$ and stiffness$^*$ $k x_{\rm m}^{2}/E_{\rm B}$. Colored curves: geodesics of the friction for equal initial and final trap stiffness ($k_{\rm i} = k_{\rm f}$). Color heatmap: absolute product of control-parameter speeds $\dot{\lambda}_{j} = \md \lambda_{j} / \md t$. In the friction, the subscript ${\rm c}$ refers to the trap center and ${\rm s}$ to the stiffness components of the matrix. The positive and negative components of the off-diagonal entry $\zeta_{\rm c,s}$ are respectively denoted by $\zeta_{\rm c,s}^{+}\equiv {\rm max}(\zeta_{\rm c,s},0) $ (b) and $\zeta_{\rm c,s}^{-}\equiv {\rm max}(-\zeta_{\rm c,s},0)$ (c). The stars in (a-d) denote the STEP for initial and final stiffness $k x_{\rm m}^{2}/E_{\rm B}=1$. (e,f) STEP (purple dotted) as a function of protocol duration, with the corresponding slow (blue dashed) and interpolated protocols (green dot-dashed). An asterisk denotes a scaled (dimensionless) quantity, with the velocities scaled by the average speed $|\dot{\lambda}_{j}\dot{\lambda}_{\ell}|^* \equiv |\dot{\lambda}_{j}\dot{\lambda}_{\ell}|/(\overline{|\dot{\lambda}_{j}|}~\overline{|\dot{\lambda}_{\ell}|})$, friction as $\zeta_{j\ell}^* \equiv \zeta_{j\ell}\lambda_{j}\lambda_{\ell}/(\lambda_{j}^{*}\lambda_{\ell}^{*}\gamma x_{\rm m}^2)$, and protocol duration as $t/\Delta t$. This figure is adapted from Fig.~S1 of the supplemental material in Ref.~\onlinecite{Blaber2022}.} 
	\label{Geodesics}
\end{figure}

To illustrate the steps for determining the minimum-work protocol within this approximation, Fig.~\ref{Geodesics} presents the friction matrix previously computed for the example model system of driven barrier crossing~\cite{Blaber2022}. For this system the friction matrix can be directly computed from \eqref{CDF friction} and the geodesics found by numerically solving \eqref{Euler-Lagrange} with specified initial and final control parameters, as described in Refs.~\onlinecite{Rotskoff2017,louwerse2022}.

Some general properties can be observed from Fig.~\ref{Geodesics}: although the friction is positive semidefinite, the off-diagonal component can become negative; geodesics never have any discontinuities (following from the Riemannian geometry); geodesics tend to avoid or slow down in regions of high friction; and for driven barrier crossing the minimum-work protocol slows down and tightens the trap as it crosses the barrier. Slowing down and tightening the trap as the system crosses the barrier is consistent with the results derived within the strong-trap approximation (section~\ref{strong trap approximation}); however, the designed slow-control protocol lacks the jumps at the start and end of the protocol.

\subsubsection{Extensions to the slow-control approximation}\label{Extensions}

The slow-driving approximation has been generalized and extended to transitions between nonequilibrium steady states~\cite{mandal2016,zulkowski2013}, discrete control~\cite{Large2019}, and stochastic control~\cite{large2018}.

Previously throughout section~\ref{Parametric control}, we assumed that the system starts in equilibrium; however, this is not always the case in applications. Periodically driven machines such as ATP synthase are often driven for a sufficiently long time as to reach a nonequilibrium steady state, breaking the initial-equilibrium and near-equilibrium assumptions. Such systems may need to transition between steady states in order to increase or decrease their output in response to variable conditions (e.g., increasing/decreasing rate of ATP production). Although determining the correct definition of dissipation is more subtle for nonequilibrium steady states (see Refs.~\onlinecite{oono1998,hatano2001,speck2005,ge2009,ge2010,bertini2012,bertini2013,bertini2015} for detailed discussion), the slow-protocol approximation has been generalized to slow transitions between nonequilibrium steady states making use of a near-steady state approximation~\cite{mandal2016,zulkowski2013}. This approximation has an analogous form to the friction-tensor approximation and can be used to determine minimum-dissipation protocols for slow transitions between nonequilibrium steady states.

So far we have assumed that the protocol is continuous; however, many biological and chemical systems convert free energy stored in nonequilibrium chemical-potential differences into useful work through a series of reactions involving binding/unbinding or catalysis of small molecules. These chemical reactions typically occur on timescales much faster than the protein conformational rearrangements they couple to. Therefore, these changes are effectively instantaneous, leading to discrete control protocols. Building off of quasistatic results~\cite{nulton1985}, it has been shown that the linear-response approximation can be applied to discretely driven systems to yield an approximation analogous to the friction tensor, which can be used to determine minimum-dissipation protocols~\cite{Large2019}.

The deterministic control considered in all other sections is a reasonable assumption for single-molecule experiments; however, autonomous systems such as ATP synthase \emph{in vivo} are driven not by a deterministic controller, but by other stochastic systems. Towards a description of fully autonomous machines, several advances have been made extending the friction-tensor framework to stochastic control~\cite{large2018} and autonomous thermodynamic cycles~\cite{bryant2020}. A detailed discussion of protocol optimization for stochastic control, as well as relations to other bounds on dissipation~\cite{machta2015} appears in Ref.~\onlinecite{large2018}.

\subsubsection{Higher-order moments and corrections}
\label{Next-order corrections and higher-order moments}

For moderate-to-fast driving, the slow-control approximation is not sufficient to accurately determine the minimum-dissipation protocol. The slow-control approximation can be generalized to higher-order moments of the work distribution and to account for the next-order correction to the excess work.\cite{Blaber2020Skewed} This is particularly useful for nonequilibrium free-energy estimation (section~\ref{Free energy estimation}).

In Ref.~\onlinecite{Blaber2020Skewed}, the higher-order corrections are derived for the first moment
\begin{align}
	\langle W_{\rm ex} \rangle_{\Lambda} &\approx \int_{0}^{\Delta t} \md t ~ \left\{\zeta_{j\ell}[\boldsymbol{\lambda}(t)]+ \frac{1}{4} \zeta^{(2)}_{j\ell m}[\boldsymbol{\lambda}(t)]\dot{\lambda}_{m}(t)\right\}\dot{\lambda}_{j}(t)\dot{\lambda}_{\ell}(t) \label{Mean_work} \ ,
\end{align}
second moment
\begin{align}
	\langle W_{\rm ex}^2 \rangle_{\Lambda} &\approx \frac{2}{\beta}\int_{0}^{\Delta t} \md t ~\left\{\zeta_{j\ell}[\boldsymbol{\lambda}(t)] + \frac{3}{4}\zeta^{(2)}_{j\ell m}[\boldsymbol{\lambda}(t)]\dot{\lambda}_{m}(t)\right\}\dot{\lambda}_{j}(t)\dot{\lambda}_{\ell}(t) \ . \label{Variance friction approximation}
\end{align}
and $n^{\rm th}$ moment
\begin{align}
\label{Higher order moment friction approximation}
\beta^{n-1} \langle W_{\rm ex}^{n} \rangle_{\Lambda} \approx \int_{0}^{\Delta t} \md t ~\left\{n\mathcal{C}_{\nu_{1}\cdots\nu_{n}}^{(n-1)}[\boldsymbol{\lambda}(t),t] + \frac{n+1}{2}\mathcal{C}_{\nu_{1}\cdots\nu_{n+1}}^{(n)}[\boldsymbol{\lambda}(t),t]\dot{\lambda}_{\nu_{n+1}}(t)\right\}\prod_{j=1}^{n}\dot{\lambda}_{\nu_{j}}(t) \ .
\end{align}
We have defined the rank-three tensor
\begin{align}
\label{rank 3 friction}
\zeta_{j\ell m}^{(2)}[\boldsymbol{\lambda}(t)]  \equiv -\beta^2\int_{0}^{\infty}\md t''\int_{0}^{\infty}\md t'''\left\langle\delta f_{j}(0)\delta f_{\ell}(t'')\delta f_{m}(t''')\right\rangle_{\boldsymbol{\lambda}(t)}  \ ,
\end{align}
and the integral $n$-time covariance functions
\begin{align}
\label{rank n friction}
\mathcal{C}_{\nu_{1}\cdots\nu_{n}}^{(n-1)}&[\boldsymbol{\lambda}(t),t] \equiv (-\beta)^{n}\prod_{j=2}^{n}\int_{0}^{t}\md t_{j}\left\langle\prod_{\ell=2}^{n}\delta f_{\nu_{1}}(0)\delta f_{\nu_{\ell}}(t_{\ell})\right\rangle_{\boldsymbol{\lambda}(t)} \ .
\end{align}
The rank-three tensor [Eq.~\eqref{rank 3 friction}] is referred to as the \emph{supra-Stokes'} tensor and is indexed by superscript (2), as it corresponds to the next-order correction to dissipation beyond the Stokes' friction~\eqref{Mean_work}. The higher-order moments of the excess work are higher order in control-parameter velocity and are therefore smaller for slow protocols.

The most notable application of this approximation is to free-energy estimation (section~\ref{Free energy estimation}), where the supra-Stokes' tensor determines the leading-order contribution to the bias of bidirectional estimators and can be used to design protocols that minimize bias~\cite{Blaber2020Skewed}.

\subsection{Interpolated protocols}\label{Interpolated protocols}

Given theory describing minimum-dissipation control in both the slow and fast limits, a simple interpolation scheme has been developed to design protocols that reduce dissipation at all driving speeds~\cite{Blaber2021,Blaber2022}. The interpolated protocol has an initial jump $(\boldsymbol{\lambda}^{\rm STEP}-\boldsymbol{\lambda}_{\rm i})/(1+\Delta t/\tau)$ and a final jump $(\boldsymbol{\lambda}_{\rm f}-\boldsymbol{\lambda}^{\rm STEP})/(1+\Delta t/\tau)$, and follows the slow-control path between them, 
\begin{align}
    \boldsymbol{\lambda}^{\rm interp}(t) = \frac{1}{1+\frac{\Delta t}{\tau}}\boldsymbol{\lambda}^{\rm STEP} + \left(1-\frac{1}{1+\frac{\Delta t}{\tau}}\right)\boldsymbol{\lambda}^{\rm slow}(t)\ ,
\end{align}
with $\tau$ the crossover duration and $\boldsymbol{\lambda}^{\rm slow}(t)$ the solution to~\eqref{Euler-Lagrange}. This guarantees that the protocol approaches the minimum-dissipation protocol in both the fast and slow limits.

\section{Free-Energy Estimation}\label{Free energy estimation}

Computational and experimental measurements of free-energy differences are essential to the determination of stable equilibrium phases of matter, relative reaction rates, and binding affinities of chemical species,\cite{Gapsys2015} and the identification and design of novel protein-binding ligands for drug discovery.\cite{Schindler2020,Kuhn2017,Ciordia2016,Wang2015} Current methods for determining free-energy differences rely on costly experimentation, which can be reduced through screening with efficient computational techniques.\cite{Schindler2020,Aldeghi2018,Kuhn2017,Ciordia2016,Wang2015,Gapsys2015,Chodera2011,Pohorille2010} It has been shown that the precision and accuracy of standard free-energy estimators are reduced when estimated from a protocol inducing large dissipation,\cite{Gore2003,Shenfeld2009,Blaber2020Skewed} and that thermodynamic geometry can be applied to improve free-energy estimates.\cite{Shenfeld2009,Minh2019,Pham2011,Pham2012,Park2014,Blaber2020Skewed}

Free-energy differences are often estimated by measuring the work incurred during a parametric control protocol that drives the system between control-parameter endpoints corresponding to target states. Unidirectional estimators determine the free-energy difference from the work done by a forward protocol driving from initial to final 
control-parameter values, while bidirectional estimators additionally use reverse protocols that drive from final to initial control-parameter values. The simplest estimator is the mean-work estimator,\cite{Gore2003} which estimates
the free-energy difference by the average work and for any non-quasistatic (finite-speed) protocol yields a biased estimate. For an unbiased estimate, the Jarzynski estimator (derived from the Jarzynski equality~\cite{Jarzynski1997}) estimates the free-energy difference from the exponentially averaged work. The mean-work and Jarzynski estimators can be used as either uni- or bidirectional estimators; however, if bidirectional data is available the maximum log-likelihood estimator is Bennett's acceptance ratio.\cite{Bennett1976} For a large number of samples, Bennett's acceptance ratio yields the minimum variance of any unbiased estimator.\cite{Shirts2003,Maragakis2006}

Near equilibrium, Taylor expanding all of these estimators for small excess work (small dissipation) gives
the mean-work estimator.~\cite{Blaber2020Skewed} For an equal number of samples in the forward and reverse directions, 
near equilibrium the variance of any of these estimators
is~\cite{Blaber2020Skewed}
\begin{subequations}
\begin{align}
\left\langle \left(\delta \widehat{\Delta F}\right)^2 \right\rangle& \approx \frac{\langle W_{\rm ex}^{2}\rangle_{\Lambda} +\langle W_{\rm ex}^{2}\rangle_{{\Lambda^{\dagger}}}}{2N} \label{Variance to second moment} \\
& \approx \frac{\langle W_{\rm ex}\rangle_{\Lambda} +\langle W_{\rm ex}\rangle_{{\Lambda^{\dagger}}}}{N} \ ,
\label{Variance_near_equilibrium}
\end{align}
\end{subequations}
where the second line follows from the fluctuation-dissipation relation for the work distribution.~\cite{Gore2003,Blaber2020Skewed} The bias is given by the asymmetry between forward $\Lambda$ and reverse $\Lambda^{\dagger}$ dissipation, generated from skewed work fluctuations:
\begin{align}
\left\langle \delta \widehat{\Delta F}\right\rangle &\approx \frac{\langle W_{\rm ex}\rangle_{\Lambda}-\langle W_{\rm ex}\rangle_{\Lambda^{\dagger}}}{2N}\label{BAR Bias} \ .
\end{align}
Equations~\eqref{Variance to second moment} and \eqref{BAR Bias} not only hold near equilibrium but also when only a single sample is taken in the forward and reverse directions, since then the average of a function is equal to the function of the average (e.g., $\langle e^x \rangle = e^{\langle x \rangle}$).

For a slow protocol the variance is approximated by
\begin{align}
\left\langle \left(\delta \widehat{\Delta F}\right)^2 \right\rangle \approx \frac{2}{\beta N}\int_{0}^{\Delta t}\md t \,  \zeta_{j\ell}[\boldsymbol{\lambda}(t)] \dot{\lambda}_{j}(t)\dot{\lambda}_{\ell}(t) \ , \label{BAR Variance Friction}
\end{align}
Thus the protocol designed to reduce the variance follows geodesics of $\zeta_{j\ell}$, and for one-dimensional control proceeds at velocity $\dot{\lambda} \propto \left(\zeta\right)^{-1/2}$. This, or the related force-variance metric~\cite{Shenfeld2009}, has been used to improve the precision of calculated binding potentials of mean force~\cite{Minh2019,Pham2011,Pham2012,Park2014}.

Unlike the variance, the protocol that maximizes the accuracy (minimizes bias) is different for unidirectional and bidirectional estimators. For unidirectional Jarzynski and mean-work estimators, near equilibrium the minimum-bias protocol is simply the minimum-dissipation protocol (protocol that minimizes \eqref{Mean_work}) and therefore to leading order is optimized by the same protocol that minimizes \eqref{BAR Variance Friction}. For a slow protocol, the bias from Bennett's acceptance ratio can be approximated as
\begin{align}
	\left\langle \delta \widehat{\Delta F}\right\rangle &\approx \frac{1}{4N}\int_{0}^{\Delta t}\md t \, \zeta^{(2)}_{j\ell m}[\boldsymbol{\lambda}(t)] \,\dot{\lambda}_{j}(t)\dot{\lambda}_{\ell}(t)\dot{\lambda}_{m}(t) \ , \label{BAR Bias Friction}
\end{align}
where the second line follows from \eqref{Mean_work}. The protocol designed to reduce the (magnitude of) bias thus follows geodesics of the cubic Finsler metric $\zeta^{(2)}_{j\ell m}$, simplifying for one-dimensional control to $\dot{\lambda} \propto \left(\zeta^{(2)}\right)^{-1/3}$.

\section{Comparison between control strategies}\label{Comparison between control strategies}

In this section we compare naive and designed protocols based on the methods described in the previous sections: interpolated protocols combining STEP and slow-protocol approximations (section~\ref{Interpolated protocols}), strong-trap approximation (section~\ref{strong trap approximation}), and full control (section~\ref{Exact solutions}). We again turn to the example model system of driven barrier crossing to assess similarities, differences, and performance of the designed protocols. The naive protocol serves as a baseline which the designed protocols should outperform, and full control as a bound on what parametric control could possibly achieve.

\begin{figure}
	\includegraphics[width=\linewidth]{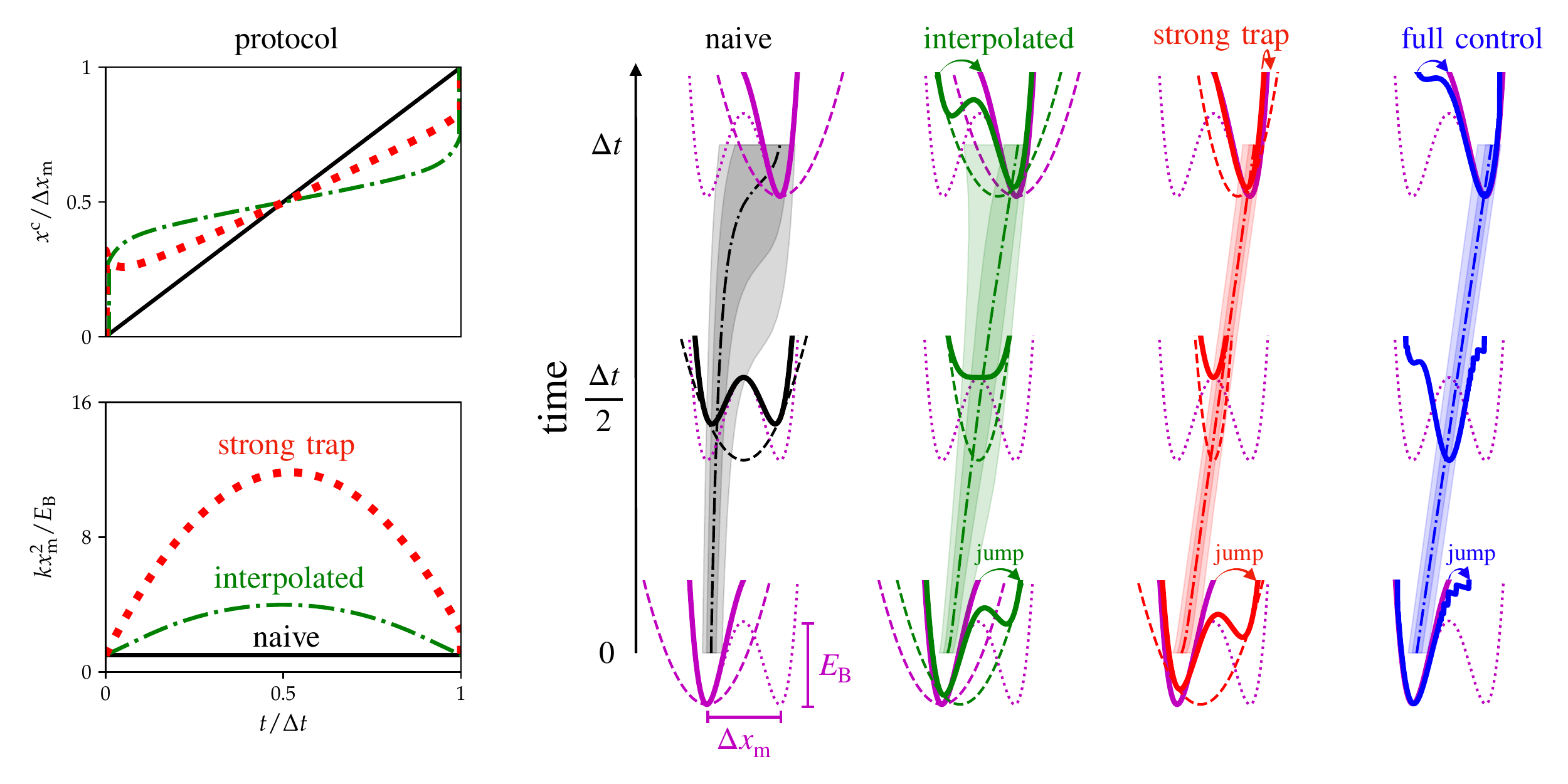}
	\caption{Time-dependent protocols for driven barrier crossing at intermediate protocol duration. Naive (black), interpolated (green), strong-trap (red), and full-control (blue) protocols. Snapshots of the total (solid), static hairpin (dotted), and time-dependent trap (dashed) potential are shown for $t=0$, $\Delta t/2$, and $\Delta t$. The hairpin, initial, and final potentials are the same across protocols (purple). Dash-dotted curves: median positions during corresponding protocol. Shading: 9\%, 25\%, 75\%, and 91\% quantiles, which are approximately evenly spaced for a Gaussian distribution. Barrier height is $E_{\rm B} = 4\beta^{-1}$, initial and final trap stiffnesses are $k_{\rm i} = k_{\rm f} = 4 \beta^{-1}/x_{\rm m}^2$, and protocol duration is $\tau_{\rm D}$ for diffusion time $\tau_{\rm D} = \Delta x_{\rm m}^2/(2D)$.} 
	\label{Protocol}
\end{figure}

Fig.~\ref{Protocol} shows the naive and designed protocols for intermediate driving speed, intermediate trap stiffness, and for fixed final control parameters. Every designed protocol has discontinuous jumps at the start and end of the protocol, and slows down and tightens the trap as it crosses the barrier. The behavior of the designed protocols can be understood in terms of the full-control solution. In one dimension the minimum-dissipation protocol linearly drives the quantiles of the probability distribution between the initial and final distributions. In the naive protocol, since it has constant stiffness, the probability distribution spreads out as it crosses the barrier due to the negative curvature of the energy landscape at the barrier. To compensate for this, the designed protocols tighten as they cross the barrier; additionally, to compensate for the changes in the gradient of the energy landscape, the designed protocols slow down as they cross the barrier.

\begin{figure}
	\includegraphics[width=\linewidth]{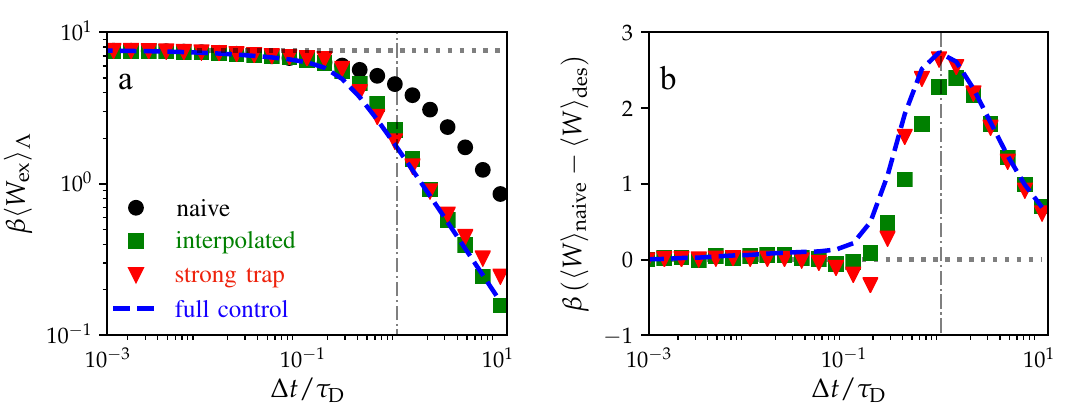}
	\caption{Performance of the naive (black circles), interpolated (green squares), strong-trap (red triangles), and full-control (blue dashed) protocols. (a) Excess work $\langle W_{\rm ex} \rangle_{\boldsymbol{\Lambda}}$ and (b) work difference $\langle W \rangle_{\rm naive} - \langle W\rangle_{\rm des}$ between designed and naive protocols, all as functions of protocol duration $\Delta t/\tau_{\rm D}$ scaled by diffusion time $\tau_{\rm D}$. Error bars representing bootstrap-resampled 95\% confidence intervals are smaller than the markers.}
	\label{Protocol_work_distance}
\end{figure}

Fig.~\ref{Protocol_work_distance} shows the excess work of the designed protocols compared to the naive protocol. For long duration (slow protocol) all of the designed protocols significantly outperform the naive, with the difference between the minimum dissipation possible from full control indistinguishable from the dissipation from the interpolated protocol in this limit. While the approximations made in the interpolated protocol become exact in the long-duration limit, the same is not true for the strong-trap approximation. As a result, the strong-control protocol has slightly higher dissipation in the long-duration limit, but would achieve the minimal dissipation in the limit of high trap stiffness. Furthermore, for intermediate protocol duration ($\Delta t \sim \tau_{\rm D}$), the strong control performs the best of the approximations since the approximation does not explicitly depend on the protocol duration. For short duration (fast protocols), all the designed protocols perform similarly well.

In summary, the designed protocols perform similarly well, so it seems reasonable to choose the control strategy which is simplest to determine, provided the approximations/assumptions required are satisfied. Since the strong-trap approximation has explicit solutions for the designed protocol, it will generally be the simplest to determine; however, it is not as widely applicable as the interpolated protocols.

\section{Perspective and outlook}\label{Perspective and outlook}

We have reviewed optimal control in stochastic thermodynamics, from full control (section~\ref{Full control}) to parametric control (section~\ref{Parametric control}). General methods are known for determining minimum-dissipation protocols for parametric control ranging from strong (section~\ref{strong trap approximation}) to weak (section~\ref{Linear response}) and fast (section~\ref{Fast control}) to slow (section~\ref{Slow control}). These approximations fill out the four limits of parametric control (Fig.~\ref{Parametric_diagram}) and can be combined to design protocols that reduce dissipation at any driving speed (section~\ref{Interpolated protocols}). These designed protocols reproduce key features determined by exact solutions for Gaussian distributions and quadratic trapping potentials, such as control-parameter jumps at the start and end of the protocol (section~\ref{Fast control}) and linear driving of the distribution quantiles between specified endpoints (section~\ref{Exact solutions}).

For the model system of driven barrier crossing (section~\ref{Model Systems}), we compare the interpolated, strong-control, and full-control solutions (section~\ref{Comparison between control strategies}). The designed protocols significantly outperform the naive (linear) protocol. Strong control has explicit solutions for the minimum-dissipation protocol, making it the simplest approximation to use; however, it is only applicable to overdamped dynamics with strong trapping potentials. 

Figure~\ref{Parametric_diagram} shows the limits in which we have solutions for the minimum-dissipation protocol for parametric control. The linear-response and slow-driving approximations have been applied to several different types of systems and control (section~\ref{Extensions}). A promising area of future study would be to explore if extensions and generalizations can be made to strong and fast control. { Indeed, it has recently been shown that the fast-protocol approximation can be extended to quantum systems, classical Hamiltonian dynamics, and optimization of the variance of the work distribution.\cite{rolandi2022}}

Another extension to consider is to allow for position-dependent diffusivity, which generically arises when a high-dimensional system is represented in a lower-dimensional space.\cite{best2010} For example, DNA hairpin experiments and molecular dynamics simulations take place in three spatial dimensions but are often represented as one-dimensional diffusive processes.\cite{hummer2005,neupane2016,neupane2017} This requires averaging out the behavior in the other two dimensions, and any inhomogeneity in these dimensions will result in position-dependent effective diffusivity in the one-dimensional representation. Measured diffusivities in molecular dynamics simulations often vary with position~\cite{hummer2005}, and although accurate measurements remain a technical challenge, some hairpin experiments report a position-dependent diffusivity.\cite{foster2018} Position-dependent diffusivity can alter the kinetic transition state of protein folding~\cite{chahine2007}, which will impact the design of minimum-dissipation protocols. Therefore, position-dependent diffusivity may be an important consideration in some applications. { The minimum-dissipation protocol under full control in inhomogeneous environments (e.g., position-dependent diffusivity) has been treated in detail in Ref.~\citenum{bo2013}.}

An important aspect of designing protocols to minimize dissipation in both experiments and theory is the choice of control parameters and number of control parameters. For the model system of driven barrier crossing, it has been shown that designed protocols with control over both trap center and stiffness (two-parameter control) significantly reduces dissipation compared to designed protocols that can only adjust the trap center (single-parameter control).~\cite{Blaber2022} However, adding even more control parameters does not significantly reduce dissipation any further, because this system is well approximated as a one-dimensional Gaussian, for which the full-control solution only requires two parameters to adjust the mean and variance (section~\ref{Exact solutions}). Although this phenomenon is well explained for one-dimensional overdamped systems, considerably less is known more generally. For example, Ref.~\onlinecite{louwerse2022} compared one-, two-, and four-parameter control of the Ising model and found significant qualitative and quantitative differences between the designed protocols. Since the full-control solution for this system is not yet known, we cannot explain this phenomenon the same way we did for driven barrier crossings. Beyond simply the number of control parameters, it remains an open question as to which control parameters are the most important when designing protocols to minimize dissipation. The choice of control parameters and number of control parameters will be important for experimental and computational applications, such as free-energy estimation.

Leveraging optimal-transport theory appears to be a promising approach towards a deeper understanding of optimal control in stochastic thermodynamics. Full-control solutions based on optimal-transport theory were initially only applicable to continuous classical systems with overdamped dynamics~\cite{Aurell2011,nakazato2021}; however, recent studies have begun to expand their range of applicability to discrete-state and quantum systems~\cite{dechant2022,dechant2022geometric,yoshimura2022,zhong2022,van2022,van2022Topological}.

The full-control solutions give an idealized view of optimal control and will always outperform parametric control, but can be used as the ideal solution that parametric control can strive towards and draw insight from. For example, linearly driving the quantiles between the initial and final endpoints is the minimum-dissipation protocol for a one-dimensional system under full control; this is reasonably well reproduced by parametric control of driven barrier crossing (section~\ref{Comparison between control strategies}). Furthermore, it has recently been shown that optimal-transport theory can be leveraged to design minimum-dissipation protocols under parametric control~\cite{zhong2022}, which is a promising new technique for determining exact minimum-dissipation protocols at any driving speed or strength of driving.

In this review we focused on protocols which reduce the average work or entropy production; however, higher-order moments (e.g., variance or skewness), or individual trajectory optimization~\cite{proesmans2022} may also be of interest for strongly fluctuating systems. Ref.~\onlinecite{Solon2018} compared for a breathing harmonic trap the protocols that minimize work variance with those that minimize average work, finding that minimum-average-work and minimum-work-variance protocols qualitatively differ for intermediate-to-fast driving speeds. In contrast, for slow near-equilibrium control, the same protocols minimize average work and work variance (section~\ref{Next-order corrections and higher-order moments}), so a more complete description of minimum-variance protocols far from equilibrium is desirable.

This review has focused on systems driven by external control parameters, which is adequate for describing the experimental systems discussed in section~\ref{Model Systems}; however, this does not accurately describe autonomous machines. For example, ATP synthase \emph{in vivo} is driven by a proton gradient across the mitochondrial membrane which drives the coupled ${\rm F}_{\rm o}$ and ${\rm F}_{1}$ components. In this system, coupling between the components means that none of the components can be treated as external, and thus it is not obvious how the present discussion of optimal control applies to such autonomous machines. Some first steps towards the description of autonomous molecular machines is discussed in section~\ref{Extensions}; however, more work is required towards a full description of efficient autonomous machines.\cite{ehrich2022}

The majority of the studies on optimal control in stochastic thermodynamics have focused on theoretical understanding and simple toy models. As we have shown in this review, there is a deep understanding of minimum-dissipation protocols at both slow and fast driving speed and both weak and strong driving strength (Fig.~\ref{Parametric_diagram}). It would be interesting to see how these results apply to real physical systems and if they are able to achieve the promising results predicted by theory. The two most straightforward applications are to relatively simple experimental model systems (section~\ref{Model Systems}) in an analogous fashion to Ref.~\onlinecite{Tafoya2019}, and to free-energy estimation as discussed in section~\ref{Free energy estimation}, in a similar fashion to Refs.~\onlinecite{Minh2019,Pham2011,Pham2012,Park2014}.

\section*{Acknowledgments}
We thank Jannik Ehrich and Matt Leighton (SFU Physics) and Miranda Louwerse (SFU Chemistry) for feedback on the manuscript. 
This work was supported by an SFU Graduate Deans Entrance Scholarship (SB), an NSERC Discovery Grant and Discovery Accelerator Supplement (DAS), and a Tier-II Canada Research Chair (DAS), and was enabled in part by support provided by WestGrid (\url{westgrid.ca}) and the Digital Research Alliance of Canada (\url{alliancecan.ca}).


\begin{thebibliography}{160}
	\expandafter\ifx\csname natexlab\endcsname\relax\def\natexlab#1{#1}\fi
	\expandafter\ifx\csname bibnamefont\endcsname\relax
	\def\bibnamefont#1{#1}\fi
	\expandafter\ifx\csname bibfnamefont\endcsname\relax
	\def\bibfnamefont#1{#1}\fi
	\expandafter\ifx\csname citenamefont\endcsname\relax
	\def\citenamefont#1{#1}\fi
	\expandafter\ifx\csname url\endcsname\relax
	\def\url#1{\texttt{#1}}\fi
	\expandafter\ifx\csname urlprefix\endcsname\relax\def\urlprefix{URL }\fi
	\providecommand{\bibinfo}[2]{#2}
	\providecommand{\eprint}[2][]{\url{#2}}
	
	\bibitem[{\citenamefont{Andresen et~al.}(1977)\citenamefont{Andresen, Berry,
			Nitzan, and Salamon}}]{andresen1977}
	\bibinfo{author}{\bibfnamefont{B.}~\bibnamefont{Andresen}},
	\bibinfo{author}{\bibfnamefont{R.~S.} \bibnamefont{Berry}},
	\bibinfo{author}{\bibfnamefont{A.}~\bibnamefont{Nitzan}}, \bibnamefont{and}
	\bibinfo{author}{\bibfnamefont{P.}~\bibnamefont{Salamon}},
	\bibinfo{journal}{Phys. Rev. A} \textbf{\bibinfo{volume}{15}},
	\bibinfo{pages}{2086} (\bibinfo{year}{1977}).
	
	\bibitem[{\citenamefont{Salamon et~al.}(1977)\citenamefont{Salamon, Andresen,
			and Berry}}]{salamon1977}
	\bibinfo{author}{\bibfnamefont{P.}~\bibnamefont{Salamon}},
	\bibinfo{author}{\bibfnamefont{B.}~\bibnamefont{Andresen}}, \bibnamefont{and}
	\bibinfo{author}{\bibfnamefont{R.~S.} \bibnamefont{Berry}},
	\bibinfo{journal}{Phys. Rev. A} \textbf{\bibinfo{volume}{15}},
	\bibinfo{pages}{2094} (\bibinfo{year}{1977}).
	
	\bibitem[{\citenamefont{Andresen}(2011)}]{andresen2011}
	\bibinfo{author}{\bibfnamefont{B.}~\bibnamefont{Andresen}},
	\bibinfo{journal}{Angew. Chem. Int. Ed.} \textbf{\bibinfo{volume}{50}},
	\bibinfo{pages}{2690} (\bibinfo{year}{2011}).
	
	\bibitem[{\citenamefont{Band et~al.}(1982)\citenamefont{Band, Kafri, and
			Salamon}}]{band1982}
	\bibinfo{author}{\bibfnamefont{Y.~B.} \bibnamefont{Band}},
	\bibinfo{author}{\bibfnamefont{O.}~\bibnamefont{Kafri}}, \bibnamefont{and}
	\bibinfo{author}{\bibfnamefont{P.}~\bibnamefont{Salamon}},
	\bibinfo{journal}{J. Appl. Phys.} \textbf{\bibinfo{volume}{53}},
	\bibinfo{pages}{8} (\bibinfo{year}{1982}).
	
	\bibitem[{\citenamefont{Weinhold}(1975)}]{Weinhold1975}
	\bibinfo{author}{\bibfnamefont{F.}~\bibnamefont{Weinhold}},
	\bibinfo{journal}{J. Chem. Phys.} \textbf{\bibinfo{volume}{63}},
	\bibinfo{pages}{2479} (\bibinfo{year}{1975}).
	
	\bibitem[{\citenamefont{Ruppeiner}(1979)}]{ruppeiner1979}
	\bibinfo{author}{\bibfnamefont{G.}~\bibnamefont{Ruppeiner}},
	\bibinfo{journal}{Phys. Rev. A} \textbf{\bibinfo{volume}{20}},
	\bibinfo{pages}{1608} (\bibinfo{year}{1979}).
	
	\bibitem[{\citenamefont{Crooks}(2007)}]{Crooks2007}
	\bibinfo{author}{\bibfnamefont{G.~E.} \bibnamefont{Crooks}},
	\bibinfo{journal}{Phys. Rev. Lett.} \textbf{\bibinfo{volume}{99}},
	\bibinfo{pages}{100602} (\bibinfo{year}{2007}).
	
	\bibitem[{\citenamefont{Salamon and Berry}(1983)}]{salamon1983}
	\bibinfo{author}{\bibfnamefont{P.}~\bibnamefont{Salamon}} \bibnamefont{and}
	\bibinfo{author}{\bibfnamefont{R.~S.} \bibnamefont{Berry}},
	\bibinfo{journal}{Phys. Rev. Lett.} \textbf{\bibinfo{volume}{51}},
	\bibinfo{pages}{1127} (\bibinfo{year}{1983}).
	
	\bibitem[{\citenamefont{Ashkin}(1970)}]{ashkin1970}
	\bibinfo{author}{\bibfnamefont{A.}~\bibnamefont{Ashkin}},
	\bibinfo{journal}{Phys. Rev. Lett.} \textbf{\bibinfo{volume}{24}},
	\bibinfo{pages}{156} (\bibinfo{year}{1970}).
	
	\bibitem[{\citenamefont{Bustamante et~al.}(2021)\citenamefont{Bustamante,
			Chemla, Liu, and Wang}}]{bustamante2021}
	\bibinfo{author}{\bibfnamefont{C.~J.} \bibnamefont{Bustamante}},
	\bibinfo{author}{\bibfnamefont{Y.~R.} \bibnamefont{Chemla}},
	\bibinfo{author}{\bibfnamefont{S.}~\bibnamefont{Liu}}, \bibnamefont{and}
	\bibinfo{author}{\bibfnamefont{M.~D.} \bibnamefont{Wang}},
	\bibinfo{journal}{Nat. Rev. Methods Primers} \textbf{\bibinfo{volume}{1}},
	\bibinfo{pages}{1} (\bibinfo{year}{2021}).
	
	\bibitem[{\citenamefont{Polimeno et~al.}(2018)\citenamefont{Polimeno, Magazzu,
			Iati, Patti, Saija, Boschi, Donato, Gucciardi, Jones, Volpe
			et~al.}}]{polimeno2018}
	\bibinfo{author}{\bibfnamefont{P.}~\bibnamefont{Polimeno}},
	\bibinfo{author}{\bibfnamefont{A.}~\bibnamefont{Magazzu}},
	\bibinfo{author}{\bibfnamefont{M.~A.} \bibnamefont{Iati}},
	\bibinfo{author}{\bibfnamefont{F.}~\bibnamefont{Patti}},
	\bibinfo{author}{\bibfnamefont{R.}~\bibnamefont{Saija}},
	\bibinfo{author}{\bibfnamefont{C.~D.~E.} \bibnamefont{Boschi}},
	\bibinfo{author}{\bibfnamefont{M.~G.} \bibnamefont{Donato}},
	\bibinfo{author}{\bibfnamefont{P.~G.} \bibnamefont{Gucciardi}},
	\bibinfo{author}{\bibfnamefont{P.~H.} \bibnamefont{Jones}},
	\bibinfo{author}{\bibfnamefont{G.}~\bibnamefont{Volpe}},
	\bibnamefont{et~al.}, \bibinfo{journal}{J. Quant. Spectrosc. Radiat. Transf.}
	\textbf{\bibinfo{volume}{218}}, \bibinfo{pages}{131} (\bibinfo{year}{2018}).
	
	\bibitem[{\citenamefont{Moffitt et~al.}(2008)\citenamefont{Moffitt, Chemla,
			Smith, and Bustamante}}]{moffitt2008}
	\bibinfo{author}{\bibfnamefont{J.~R.} \bibnamefont{Moffitt}},
	\bibinfo{author}{\bibfnamefont{Y.~R.} \bibnamefont{Chemla}},
	\bibinfo{author}{\bibfnamefont{S.~B.} \bibnamefont{Smith}}, \bibnamefont{and}
	\bibinfo{author}{\bibfnamefont{C.}~\bibnamefont{Bustamante}},
	\bibinfo{journal}{Annu. Rev. Biochem.} \textbf{\bibinfo{volume}{77}},
	\bibinfo{pages}{205} (\bibinfo{year}{2008}).
	
	\bibitem[{\citenamefont{Liphardt et~al.}(2002)\citenamefont{Liphardt, Dumont,
			Smith, Tinoco, and Bustamante}}]{liphardt2002}
	\bibinfo{author}{\bibfnamefont{J.}~\bibnamefont{Liphardt}},
	\bibinfo{author}{\bibfnamefont{S.}~\bibnamefont{Dumont}},
	\bibinfo{author}{\bibfnamefont{S.~B.} \bibnamefont{Smith}},
	\bibinfo{author}{\bibfnamefont{I.}~\bibnamefont{Tinoco}}, \bibnamefont{and}
	\bibinfo{author}{\bibfnamefont{C.}~\bibnamefont{Bustamante}},
	\bibinfo{journal}{Science} \textbf{\bibinfo{volume}{296}},
	\bibinfo{pages}{1832} (\bibinfo{year}{2002}).
	
	\bibitem[{\citenamefont{Collin et~al.}(2005)\citenamefont{Collin, Ritort,
			Jarzynski, Smith, Tinoco, and Bustamante}}]{collin2005}
	\bibinfo{author}{\bibfnamefont{D.}~\bibnamefont{Collin}},
	\bibinfo{author}{\bibfnamefont{F.}~\bibnamefont{Ritort}},
	\bibinfo{author}{\bibfnamefont{C.}~\bibnamefont{Jarzynski}},
	\bibinfo{author}{\bibfnamefont{S.~B.} \bibnamefont{Smith}},
	\bibinfo{author}{\bibfnamefont{I.}~\bibnamefont{Tinoco}}, \bibnamefont{and}
	\bibinfo{author}{\bibfnamefont{C.}~\bibnamefont{Bustamante}},
	\bibinfo{journal}{Nature} \textbf{\bibinfo{volume}{437}},
	\bibinfo{pages}{231} (\bibinfo{year}{2005}).
	
	\bibitem[{\citenamefont{Bustamante et~al.}(2000)\citenamefont{Bustamante,
			Smith, Liphardt, and Smith}}]{bustamante2000}
	\bibinfo{author}{\bibfnamefont{C.}~\bibnamefont{Bustamante}},
	\bibinfo{author}{\bibfnamefont{S.~B.} \bibnamefont{Smith}},
	\bibinfo{author}{\bibfnamefont{J.}~\bibnamefont{Liphardt}}, \bibnamefont{and}
	\bibinfo{author}{\bibfnamefont{D.}~\bibnamefont{Smith}},
	\bibinfo{journal}{Curr. Opin. Struct. Biol.} \textbf{\bibinfo{volume}{10}},
	\bibinfo{pages}{279} (\bibinfo{year}{2000}).
	
	\bibitem[{\citenamefont{Bustamante et~al.}(2003)\citenamefont{Bustamante,
			Bryant, and Smith}}]{bustamante2003}
	\bibinfo{author}{\bibfnamefont{C.}~\bibnamefont{Bustamante}},
	\bibinfo{author}{\bibfnamefont{Z.}~\bibnamefont{Bryant}}, \bibnamefont{and}
	\bibinfo{author}{\bibfnamefont{S.~B.} \bibnamefont{Smith}},
	\bibinfo{journal}{Nature} \textbf{\bibinfo{volume}{421}},
	\bibinfo{pages}{423} (\bibinfo{year}{2003}).
	
	\bibitem[{\citenamefont{Woodside et~al.}(2006)\citenamefont{Woodside,
			Behnke-Parks, Larizadeh, Travers, Herschlag, and Block}}]{woodside2006}
	\bibinfo{author}{\bibfnamefont{M.~T.} \bibnamefont{Woodside}},
	\bibinfo{author}{\bibfnamefont{W.~M.} \bibnamefont{Behnke-Parks}},
	\bibinfo{author}{\bibfnamefont{K.}~\bibnamefont{Larizadeh}},
	\bibinfo{author}{\bibfnamefont{K.}~\bibnamefont{Travers}},
	\bibinfo{author}{\bibfnamefont{D.}~\bibnamefont{Herschlag}},
	\bibnamefont{and} \bibinfo{author}{\bibfnamefont{S.~M.} \bibnamefont{Block}},
	\bibinfo{journal}{Proc. Natl. Acad. Sci. U.S.A}
	\textbf{\bibinfo{volume}{103}}, \bibinfo{pages}{6190} (\bibinfo{year}{2006}).
	
	\bibitem[{\citenamefont{Neupane et~al.}(2017)\citenamefont{Neupane, Wang, and
			Woodside}}]{neupane2017}
	\bibinfo{author}{\bibfnamefont{K.}~\bibnamefont{Neupane}},
	\bibinfo{author}{\bibfnamefont{F.}~\bibnamefont{Wang}}, \bibnamefont{and}
	\bibinfo{author}{\bibfnamefont{M.~T.} \bibnamefont{Woodside}},
	\bibinfo{journal}{Proc. Natl. Acad. Sci. U.S.A}
	\textbf{\bibinfo{volume}{114}}, \bibinfo{pages}{1329} (\bibinfo{year}{2017}).
	
	\bibitem[{\citenamefont{Unksov et~al.}(2022)\citenamefont{Unksov, Korosec,
			Surendiran, Verardo, Lyttleton, Forde, and Linke}}]{unksov2022}
	\bibinfo{author}{\bibfnamefont{I.~N.} \bibnamefont{Unksov}},
	\bibinfo{author}{\bibfnamefont{C.~S.} \bibnamefont{Korosec}},
	\bibinfo{author}{\bibfnamefont{P.}~\bibnamefont{Surendiran}},
	\bibinfo{author}{\bibfnamefont{D.}~\bibnamefont{Verardo}},
	\bibinfo{author}{\bibfnamefont{R.}~\bibnamefont{Lyttleton}},
	\bibinfo{author}{\bibfnamefont{N.~R.} \bibnamefont{Forde}}, \bibnamefont{and}
	\bibinfo{author}{\bibfnamefont{H.}~\bibnamefont{Linke}},
	\bibinfo{journal}{ACS nanoscience Au} \textbf{\bibinfo{volume}{2}},
	\bibinfo{pages}{140} (\bibinfo{year}{2022}).
	
	\bibitem[{\citenamefont{Toyabe et~al.}(2011)\citenamefont{Toyabe,
			Watanabe-Nakayama, Okamoto, Kudo, and Muneyuki}}]{Toyabe2011}
	\bibinfo{author}{\bibfnamefont{S.}~\bibnamefont{Toyabe}},
	\bibinfo{author}{\bibfnamefont{T.}~\bibnamefont{Watanabe-Nakayama}},
	\bibinfo{author}{\bibfnamefont{T.}~\bibnamefont{Okamoto}},
	\bibinfo{author}{\bibfnamefont{S.}~\bibnamefont{Kudo}}, \bibnamefont{and}
	\bibinfo{author}{\bibfnamefont{E.}~\bibnamefont{Muneyuki}},
	\bibinfo{journal}{Proc. Natl. Acad. Sci. U.S.A}
	\textbf{\bibinfo{volume}{108}}, \bibinfo{pages}{17951}
	(\bibinfo{year}{2011}).
	
	\bibitem[{\citenamefont{Toyabe et~al.}(2012)\citenamefont{Toyabe, Ueno, and
			Muneyuki}}]{toyabe2012}
	\bibinfo{author}{\bibfnamefont{S.}~\bibnamefont{Toyabe}},
	\bibinfo{author}{\bibfnamefont{H.}~\bibnamefont{Ueno}}, \bibnamefont{and}
	\bibinfo{author}{\bibfnamefont{E.}~\bibnamefont{Muneyuki}},
	\bibinfo{journal}{Europhys. Lett.} \textbf{\bibinfo{volume}{97}},
	\bibinfo{pages}{40004} (\bibinfo{year}{2012}).
	
	\bibitem[{\citenamefont{Kawaguchi et~al.}(2014)\citenamefont{Kawaguchi, Sasa,
			and Sagawa}}]{kawaguchi2014}
	\bibinfo{author}{\bibfnamefont{K.}~\bibnamefont{Kawaguchi}},
	\bibinfo{author}{\bibfnamefont{S.-i.} \bibnamefont{Sasa}}, \bibnamefont{and}
	\bibinfo{author}{\bibfnamefont{T.}~\bibnamefont{Sagawa}},
	\bibinfo{journal}{Biophys. J.} \textbf{\bibinfo{volume}{106}},
	\bibinfo{pages}{2450} (\bibinfo{year}{2014}).
	
	\bibitem[{\citenamefont{Svoboda et~al.}(1993)\citenamefont{Svoboda, Schmidt,
			Schnapp, and Block}}]{svoboda1993}
	\bibinfo{author}{\bibfnamefont{K.}~\bibnamefont{Svoboda}},
	\bibinfo{author}{\bibfnamefont{C.~F.} \bibnamefont{Schmidt}},
	\bibinfo{author}{\bibfnamefont{B.~J.} \bibnamefont{Schnapp}},
	\bibnamefont{and} \bibinfo{author}{\bibfnamefont{S.~M.} \bibnamefont{Block}},
	\bibinfo{journal}{Nature} \textbf{\bibinfo{volume}{365}},
	\bibinfo{pages}{721} (\bibinfo{year}{1993}).
	
	\bibitem[{\citenamefont{Svoboda and Block}(1994)}]{svoboda1994}
	\bibinfo{author}{\bibfnamefont{K.}~\bibnamefont{Svoboda}} \bibnamefont{and}
	\bibinfo{author}{\bibfnamefont{S.~M.} \bibnamefont{Block}},
	\bibinfo{journal}{Cell} \textbf{\bibinfo{volume}{77}}, \bibinfo{pages}{773}
	(\bibinfo{year}{1994}).
	
	\bibitem[{\citenamefont{Kojima et~al.}(1997)\citenamefont{Kojima, Muto,
			Higuchi, and Yanagida}}]{kojima1997}
	\bibinfo{author}{\bibfnamefont{H.}~\bibnamefont{Kojima}},
	\bibinfo{author}{\bibfnamefont{E.}~\bibnamefont{Muto}},
	\bibinfo{author}{\bibfnamefont{H.}~\bibnamefont{Higuchi}}, \bibnamefont{and}
	\bibinfo{author}{\bibfnamefont{T.}~\bibnamefont{Yanagida}},
	\bibinfo{journal}{Biophys. J.} \textbf{\bibinfo{volume}{73}},
	\bibinfo{pages}{2012} (\bibinfo{year}{1997}).
	
	\bibitem[{\citenamefont{Hunt et~al.}(1994)\citenamefont{Hunt, Gittes, and
			Howard}}]{hunt1994}
	\bibinfo{author}{\bibfnamefont{A.~J.} \bibnamefont{Hunt}},
	\bibinfo{author}{\bibfnamefont{F.}~\bibnamefont{Gittes}}, \bibnamefont{and}
	\bibinfo{author}{\bibfnamefont{J.}~\bibnamefont{Howard}},
	\bibinfo{journal}{Biophys. J.} \textbf{\bibinfo{volume}{67}},
	\bibinfo{pages}{766} (\bibinfo{year}{1994}).
	
	\bibitem[{\citenamefont{Greenberg et~al.}(2016)\citenamefont{Greenberg, Arpağ,
			Tüzel, and Ostap}}]{greenberg2016}
	\bibinfo{author}{\bibfnamefont{M.}~\bibnamefont{Greenberg}},
	\bibinfo{author}{\bibfnamefont{G.}~\bibnamefont{Arpağ}},
	\bibinfo{author}{\bibfnamefont{E.}~\bibnamefont{Tüzel}}, \bibnamefont{and}
	\bibinfo{author}{\bibfnamefont{E.}~\bibnamefont{Ostap}},
	\bibinfo{journal}{Biophys. J.} \textbf{\bibinfo{volume}{110}},
	\bibinfo{pages}{2568} (\bibinfo{year}{2016}).
	
	\bibitem[{\citenamefont{Laakso et~al.}(2008)\citenamefont{Laakso, Lewis,
			Shuman, and Ostap}}]{Laakso2008}
	\bibinfo{author}{\bibfnamefont{J.~M.} \bibnamefont{Laakso}},
	\bibinfo{author}{\bibfnamefont{J.~H.} \bibnamefont{Lewis}},
	\bibinfo{author}{\bibfnamefont{H.}~\bibnamefont{Shuman}}, \bibnamefont{and}
	\bibinfo{author}{\bibfnamefont{E.~M.} \bibnamefont{Ostap}},
	\bibinfo{journal}{Science} \textbf{\bibinfo{volume}{321}},
	\bibinfo{pages}{133} (\bibinfo{year}{2008}).
	
	\bibitem[{\citenamefont{Norstrom et~al.}(2010)\citenamefont{Norstrom,
			Smithback, and Rock}}]{norstrom2010}
	\bibinfo{author}{\bibfnamefont{M.~F.} \bibnamefont{Norstrom}},
	\bibinfo{author}{\bibfnamefont{P.~A.} \bibnamefont{Smithback}},
	\bibnamefont{and} \bibinfo{author}{\bibfnamefont{R.~S.} \bibnamefont{Rock}},
	\bibinfo{journal}{J. Biol. Chem.} \textbf{\bibinfo{volume}{285}},
	\bibinfo{pages}{26326} (\bibinfo{year}{2010}).
	
	\bibitem[{\citenamefont{Nagy et~al.}(2013)\citenamefont{Nagy, Takagi,
			Billington, Sun, Hong, Homsher, Wang, and Sellers}}]{nagy2013}
	\bibinfo{author}{\bibfnamefont{A.}~\bibnamefont{Nagy}},
	\bibinfo{author}{\bibfnamefont{Y.}~\bibnamefont{Takagi}},
	\bibinfo{author}{\bibfnamefont{N.}~\bibnamefont{Billington}},
	\bibinfo{author}{\bibfnamefont{S.~A.} \bibnamefont{Sun}},
	\bibinfo{author}{\bibfnamefont{D.~K.} \bibnamefont{Hong}},
	\bibinfo{author}{\bibfnamefont{E.}~\bibnamefont{Homsher}},
	\bibinfo{author}{\bibfnamefont{A.}~\bibnamefont{Wang}}, \bibnamefont{and}
	\bibinfo{author}{\bibfnamefont{J.~R.} \bibnamefont{Sellers}},
	\bibinfo{journal}{J. Biol. Chem.} \textbf{\bibinfo{volume}{288}},
	\bibinfo{pages}{709} (\bibinfo{year}{2013}).
	
	\bibitem[{\citenamefont{Seifert}(2012)}]{Seifert2012}
	\bibinfo{author}{\bibfnamefont{U.}~\bibnamefont{Seifert}},
	\bibinfo{journal}{Rep. Prog. Phys.} \textbf{\bibinfo{volume}{75}},
	\bibinfo{pages}{126001} (\bibinfo{year}{2012}).
	
	\bibitem[{\citenamefont{Jarzynski}(2011)}]{Jarzynski2011}
	\bibinfo{author}{\bibfnamefont{C.}~\bibnamefont{Jarzynski}},
	\bibinfo{journal}{Annu. Rev. Condens. Matter Phys.}
	\textbf{\bibinfo{volume}{2}}, \bibinfo{pages}{329} (\bibinfo{year}{2011}).
	
	\bibitem[{\citenamefont{Brown and Sivak}(2017)}]{Brown2017}
	\bibinfo{author}{\bibfnamefont{A.~I.} \bibnamefont{Brown}} \bibnamefont{and}
	\bibinfo{author}{\bibfnamefont{D.~A.} \bibnamefont{Sivak}},
	\bibinfo{journal}{Physics in Canada} \textbf{\bibinfo{volume}{73}}
	(\bibinfo{year}{2017}).
	
	\bibitem[{\citenamefont{Brown and Sivak}(2019)}]{Brown2019}
	\bibinfo{author}{\bibfnamefont{A.~I.} \bibnamefont{Brown}} \bibnamefont{and}
	\bibinfo{author}{\bibfnamefont{D.~A.} \bibnamefont{Sivak}},
	\bibinfo{journal}{Chem. Rev.} \textbf{\bibinfo{volume}{120}},
	\bibinfo{pages}{434} (\bibinfo{year}{2019}).
	
	\bibitem[{\citenamefont{Schmiedl and Seifert}(2007)}]{Schmiedl2007}
	\bibinfo{author}{\bibfnamefont{T.}~\bibnamefont{Schmiedl}} \bibnamefont{and}
	\bibinfo{author}{\bibfnamefont{U.}~\bibnamefont{Seifert}},
	\bibinfo{journal}{Phys. Rev. Lett.} \textbf{\bibinfo{volume}{98}},
	\bibinfo{pages}{108301} (\bibinfo{year}{2007}).
	
	\bibitem[{\citenamefont{Abiuso et~al.}(2022)\citenamefont{Abiuso, Holubec,
			Anders, Ye, Cerisola, and Perarnau-Llobet}}]{abiuso2022}
	\bibinfo{author}{\bibfnamefont{P.}~\bibnamefont{Abiuso}},
	\bibinfo{author}{\bibfnamefont{V.}~\bibnamefont{Holubec}},
	\bibinfo{author}{\bibfnamefont{J.}~\bibnamefont{Anders}},
	\bibinfo{author}{\bibfnamefont{Z.}~\bibnamefont{Ye}},
	\bibinfo{author}{\bibfnamefont{F.}~\bibnamefont{Cerisola}}, \bibnamefont{and}
	\bibinfo{author}{\bibfnamefont{M.}~\bibnamefont{Perarnau-Llobet}},
	\bibinfo{journal}{J. Phys. Commun.} \textbf{\bibinfo{volume}{6}},
	\bibinfo{pages}{063001} (\bibinfo{year}{2022}).
	
	\bibitem[{\citenamefont{Blaber and
			Sivak}(2022{\natexlab{a}})}]{Blaber2022_strong}
	\bibinfo{author}{\bibfnamefont{S.}~\bibnamefont{Blaber}} \bibnamefont{and}
	\bibinfo{author}{\bibfnamefont{D.~A.} \bibnamefont{Sivak}},
	\bibinfo{journal}{Phys. Rev. E} \textbf{\bibinfo{volume}{106}},
	\bibinfo{pages}{L022103} (\bibinfo{year}{2022}{\natexlab{a}}).
	
	\bibitem[{\citenamefont{Gomez-Marin et~al.}(2008)\citenamefont{Gomez-Marin,
			Schmiedl, and Seifert}}]{Gomez2008}
	\bibinfo{author}{\bibfnamefont{A.}~\bibnamefont{Gomez-Marin}},
	\bibinfo{author}{\bibfnamefont{T.}~\bibnamefont{Schmiedl}}, \bibnamefont{and}
	\bibinfo{author}{\bibfnamefont{U.}~\bibnamefont{Seifert}},
	\bibinfo{journal}{J. Chem. Phys.} \textbf{\bibinfo{volume}{129}},
	\bibinfo{pages}{024114} (\bibinfo{year}{2008}).
	
	\bibitem[{\citenamefont{Then and Engel}(2008)}]{Then2008}
	\bibinfo{author}{\bibfnamefont{H.}~\bibnamefont{Then}} \bibnamefont{and}
	\bibinfo{author}{\bibfnamefont{A.}~\bibnamefont{Engel}},
	\bibinfo{journal}{Phys. Rev. E} \textbf{\bibinfo{volume}{77}},
	\bibinfo{pages}{041105} (\bibinfo{year}{2008}).
	
	\bibitem[{\citenamefont{Esposito
			et~al.}(2010{\natexlab{a}})\citenamefont{Esposito, Kawai, Lindenberg, and
			Van~den Broeck}}]{Esposito2010}
	\bibinfo{author}{\bibfnamefont{M.}~\bibnamefont{Esposito}},
	\bibinfo{author}{\bibfnamefont{R.}~\bibnamefont{Kawai}},
	\bibinfo{author}{\bibfnamefont{K.}~\bibnamefont{Lindenberg}},
	\bibnamefont{and} \bibinfo{author}{\bibfnamefont{C.}~\bibnamefont{Van~den
			Broeck}}, \bibinfo{journal}{Europhys. Lett.} \textbf{\bibinfo{volume}{89}},
	\bibinfo{pages}{20003} (\bibinfo{year}{2010}{\natexlab{a}}).
	
	\bibitem[{\citenamefont{Blaber et~al.}(2021)\citenamefont{Blaber, Louwerse, and
			Sivak}}]{Blaber2021}
	\bibinfo{author}{\bibfnamefont{S.}~\bibnamefont{Blaber}},
	\bibinfo{author}{\bibfnamefont{M.~D.} \bibnamefont{Louwerse}},
	\bibnamefont{and} \bibinfo{author}{\bibfnamefont{D.~A.} \bibnamefont{Sivak}},
	\bibinfo{journal}{Phys. Rev. E} \textbf{\bibinfo{volume}{104}},
	\bibinfo{pages}{L022101} (\bibinfo{year}{2021}).
	
	\bibitem[{\citenamefont{Villani}(2009)}]{villani2009}
	\bibinfo{author}{\bibfnamefont{C.}~\bibnamefont{Villani}},
	\emph{\bibinfo{title}{Optimal transport: old and new}}, vol.
	\bibinfo{volume}{338} (\bibinfo{publisher}{Springer}, \bibinfo{year}{2009}).
	
	\bibitem[{\citenamefont{Santambrogio}(2015)}]{santambrogio2015}
	\bibinfo{author}{\bibfnamefont{F.}~\bibnamefont{Santambrogio}},
	\emph{\bibinfo{title}{Optimal transport for applied mathematicians}},
	vol.~\bibinfo{volume}{55} (\bibinfo{publisher}{Springer},
	\bibinfo{year}{2015}).
	
	\bibitem[{\citenamefont{Aurell et~al.}(2011)\citenamefont{Aurell,
			Mej{\'\i}a-Monasterio, and Muratore-Ginanneschi}}]{Aurell2011}
	\bibinfo{author}{\bibfnamefont{E.}~\bibnamefont{Aurell}},
	\bibinfo{author}{\bibfnamefont{C.}~\bibnamefont{Mej{\'\i}a-Monasterio}},
	\bibnamefont{and}
	\bibinfo{author}{\bibfnamefont{P.}~\bibnamefont{Muratore-Ginanneschi}},
	\bibinfo{journal}{Phys. Rev. Lett.} \textbf{\bibinfo{volume}{106}},
	\bibinfo{pages}{250601} (\bibinfo{year}{2011}).
	
	\bibitem[{\citenamefont{Zhang}(2019)}]{Zhang2019}
	\bibinfo{author}{\bibfnamefont{Y.}~\bibnamefont{Zhang}},
	\bibinfo{journal}{Europhys. Lett.} \textbf{\bibinfo{volume}{128}},
	\bibinfo{pages}{30002} (\bibinfo{year}{2019}).
	
	\bibitem[{\citenamefont{Nakazato and Ito}(2021)}]{nakazato2021}
	\bibinfo{author}{\bibfnamefont{M.}~\bibnamefont{Nakazato}} \bibnamefont{and}
	\bibinfo{author}{\bibfnamefont{S.}~\bibnamefont{Ito}},
	\bibinfo{journal}{Phys. Rev. Res.} \textbf{\bibinfo{volume}{3}},
	\bibinfo{pages}{043093} (\bibinfo{year}{2021}).
	
	\bibitem[{\citenamefont{Dechant}(2022)}]{dechant2022}
	\bibinfo{author}{\bibfnamefont{A.}~\bibnamefont{Dechant}}, \bibinfo{journal}{J.
		Phys. A Math. Theor.} \textbf{\bibinfo{volume}{55}}, \bibinfo{pages}{094001}
	(\bibinfo{year}{2022}).
	
	\bibitem[{\citenamefont{Miangolarra et~al.}(2022)\citenamefont{Miangolarra,
			Taghvaei, Chen, and Georgiou}}]{miangolarra2022}
	\bibinfo{author}{\bibfnamefont{O.~M.} \bibnamefont{Miangolarra}},
	\bibinfo{author}{\bibfnamefont{A.}~\bibnamefont{Taghvaei}},
	\bibinfo{author}{\bibfnamefont{Y.}~\bibnamefont{Chen}}, \bibnamefont{and}
	\bibinfo{author}{\bibfnamefont{T.~T.} \bibnamefont{Georgiou}},
	\bibinfo{journal}{IEEE Contr. Syst. Lett.} \textbf{\bibinfo{volume}{6}},
	\bibinfo{pages}{3409} (\bibinfo{year}{2022}).
	
	\bibitem[{\citenamefont{Benamou and Brenier}(2000)}]{benamou2000}
	\bibinfo{author}{\bibfnamefont{J.-D.} \bibnamefont{Benamou}} \bibnamefont{and}
	\bibinfo{author}{\bibfnamefont{Y.}~\bibnamefont{Brenier}},
	\bibinfo{journal}{Numerische Mathematik} \textbf{\bibinfo{volume}{84}},
	\bibinfo{pages}{375} (\bibinfo{year}{2000}).
	
	\bibitem[{\citenamefont{Proesmans
			et~al.}(2020{\natexlab{a}})\citenamefont{Proesmans, Ehrich, and
			Bechhoefer}}]{Proesmans2020}
	\bibinfo{author}{\bibfnamefont{K.}~\bibnamefont{Proesmans}},
	\bibinfo{author}{\bibfnamefont{J.}~\bibnamefont{Ehrich}}, \bibnamefont{and}
	\bibinfo{author}{\bibfnamefont{J.}~\bibnamefont{Bechhoefer}},
	\bibinfo{journal}{Phys. Rev. Lett.} \textbf{\bibinfo{volume}{125}},
	\bibinfo{pages}{100602} (\bibinfo{year}{2020}{\natexlab{a}}).
	
	\bibitem[{\citenamefont{Proesmans
			et~al.}(2020{\natexlab{b}})\citenamefont{Proesmans, Ehrich, and
			Bechhoefer}}]{proesmans2020optimal}
	\bibinfo{author}{\bibfnamefont{K.}~\bibnamefont{Proesmans}},
	\bibinfo{author}{\bibfnamefont{J.}~\bibnamefont{Ehrich}}, \bibnamefont{and}
	\bibinfo{author}{\bibfnamefont{J.}~\bibnamefont{Bechhoefer}},
	\bibinfo{journal}{Phys. Rev. E} \textbf{\bibinfo{volume}{102}},
	\bibinfo{pages}{032105} (\bibinfo{year}{2020}{\natexlab{b}}).
	
	\bibitem[{\citenamefont{Dechant et~al.}(2022)\citenamefont{Dechant, Sasa, and
			Ito}}]{dechant2022geometric}
	\bibinfo{author}{\bibfnamefont{A.}~\bibnamefont{Dechant}},
	\bibinfo{author}{\bibfnamefont{S.-i.} \bibnamefont{Sasa}}, \bibnamefont{and}
	\bibinfo{author}{\bibfnamefont{S.}~\bibnamefont{Ito}},
	\bibinfo{journal}{Phys. Rev. Res.} \textbf{\bibinfo{volume}{4}},
	\bibinfo{pages}{L012034} (\bibinfo{year}{2022}).
	
	\bibitem[{\citenamefont{Yoshimura et~al.}(2022)\citenamefont{Yoshimura,
			Kolchinsky, Dechant, and Ito}}]{yoshimura2022}
	\bibinfo{author}{\bibfnamefont{K.}~\bibnamefont{Yoshimura}},
	\bibinfo{author}{\bibfnamefont{A.}~\bibnamefont{Kolchinsky}},
	\bibinfo{author}{\bibfnamefont{A.}~\bibnamefont{Dechant}}, \bibnamefont{and}
	\bibinfo{author}{\bibfnamefont{S.}~\bibnamefont{Ito}},
	\bibinfo{journal}{arXiv preprint arXiv:2205.15227}  (\bibinfo{year}{2022}).
	
	\bibitem[{\citenamefont{Zhong and DeWeese}(2022)}]{zhong2022}
	\bibinfo{author}{\bibfnamefont{A.}~\bibnamefont{Zhong}} \bibnamefont{and}
	\bibinfo{author}{\bibfnamefont{M.~R.} \bibnamefont{DeWeese}},
	\bibinfo{journal}{Phys. Rev. E} \textbf{\bibinfo{volume}{106}},
	\bibinfo{pages}{044135} (\bibinfo{year}{2022}).
	
	\bibitem[{\citenamefont{Van~Vu and Saito}(2023{\natexlab{a}})}]{van2022}
	\bibinfo{author}{\bibfnamefont{T.}~\bibnamefont{Van~Vu}} \bibnamefont{and}
	\bibinfo{author}{\bibfnamefont{K.}~\bibnamefont{Saito}},
	\bibinfo{journal}{Phys. Rev. X} \textbf{\bibinfo{volume}{13}},
	\bibinfo{pages}{011013} (\bibinfo{year}{2023}{\natexlab{a}}).
	
	\bibitem[{\citenamefont{Van~Vu and
			Saito}(2023{\natexlab{b}})}]{van2022Topological}
	\bibinfo{author}{\bibfnamefont{T.}~\bibnamefont{Van~Vu}} \bibnamefont{and}
	\bibinfo{author}{\bibfnamefont{K.}~\bibnamefont{Saito}},
	\bibinfo{journal}{Phys. Rev. Lett.} \textbf{\bibinfo{volume}{130}},
	\bibinfo{pages}{010402} (\bibinfo{year}{2023}{\natexlab{b}}).
	
	\bibitem[{\citenamefont{Gingrich et~al.}(2016)\citenamefont{Gingrich, Rotskoff,
			Crooks, and Geissler}}]{gingrich2016}
	\bibinfo{author}{\bibfnamefont{T.~R.} \bibnamefont{Gingrich}},
	\bibinfo{author}{\bibfnamefont{G.~M.} \bibnamefont{Rotskoff}},
	\bibinfo{author}{\bibfnamefont{G.~E.} \bibnamefont{Crooks}},
	\bibnamefont{and} \bibinfo{author}{\bibfnamefont{P.~L.}
		\bibnamefont{Geissler}}, \bibinfo{journal}{Proc. Natl. Acad. Sci. U.S.A.}
	\textbf{\bibinfo{volume}{113}}, \bibinfo{pages}{10263}
	(\bibinfo{year}{2016}).
	
	\bibitem[{\citenamefont{Engel et~al.}(2022)\citenamefont{Engel, Smith, and
			Brenner}}]{engel2022}
	\bibinfo{author}{\bibfnamefont{M.~C.} \bibnamefont{Engel}},
	\bibinfo{author}{\bibfnamefont{J.~A.} \bibnamefont{Smith}}, \bibnamefont{and}
	\bibinfo{author}{\bibfnamefont{M.~P.} \bibnamefont{Brenner}},
	\bibinfo{journal}{arXiv preprint arXiv:2201.00098}  (\bibinfo{year}{2022}).
	
	\bibitem[{\citenamefont{Bonan{\c{c}}a and Deffner}(2018)}]{bonanca2018}
	\bibinfo{author}{\bibfnamefont{M.~V.} \bibnamefont{Bonan{\c{c}}a}}
	\bibnamefont{and} \bibinfo{author}{\bibfnamefont{S.}~\bibnamefont{Deffner}},
	\bibinfo{journal}{Phys. Rev. E} \textbf{\bibinfo{volume}{98}},
	\bibinfo{pages}{042103} (\bibinfo{year}{2018}).
	
	\bibitem[{\citenamefont{Kamizaki et~al.}(2022)\citenamefont{Kamizaki,
			Bonan{\c{c}}a, and Muniz}}]{kamizaki2022}
	\bibinfo{author}{\bibfnamefont{L.~P.} \bibnamefont{Kamizaki}},
	\bibinfo{author}{\bibfnamefont{M.~V.} \bibnamefont{Bonan{\c{c}}a}},
	\bibnamefont{and} \bibinfo{author}{\bibfnamefont{S.~R.} \bibnamefont{Muniz}},
	\bibinfo{journal}{Phys. Rev. E} \textbf{\bibinfo{volume}{106}},
	\bibinfo{pages}{064123} (\bibinfo{year}{2022}).
	
	\bibitem[{\citenamefont{Sivak and Crooks}(2016)}]{Sivak2016}
	\bibinfo{author}{\bibfnamefont{D.~A.} \bibnamefont{Sivak}} \bibnamefont{and}
	\bibinfo{author}{\bibfnamefont{G.~E.} \bibnamefont{Crooks}},
	\bibinfo{journal}{Phys. Rev. E} \textbf{\bibinfo{volume}{94}},
	\bibinfo{pages}{052106} (\bibinfo{year}{2016}).
	
	\bibitem[{\citenamefont{Blaber and Sivak}(2020{\natexlab{a}})}]{Blaber2020}
	\bibinfo{author}{\bibfnamefont{S.}~\bibnamefont{Blaber}} \bibnamefont{and}
	\bibinfo{author}{\bibfnamefont{D.~A.} \bibnamefont{Sivak}},
	\bibinfo{journal}{Phys. Rev. E} \textbf{\bibinfo{volume}{101}},
	\bibinfo{pages}{022118} (\bibinfo{year}{2020}{\natexlab{a}}).
	
	\bibitem[{\citenamefont{Deffner and Bonan{\c{c}}a}(2020)}]{deffner2020}
	\bibinfo{author}{\bibfnamefont{S.}~\bibnamefont{Deffner}} \bibnamefont{and}
	\bibinfo{author}{\bibfnamefont{M.~V.} \bibnamefont{Bonan{\c{c}}a}},
	\bibinfo{journal}{Europhys. Lett.} \textbf{\bibinfo{volume}{131}},
	\bibinfo{pages}{20001} (\bibinfo{year}{2020}).
	
	\bibitem[{\citenamefont{Zulkowski et~al.}(2012)\citenamefont{Zulkowski, Sivak,
			Crooks, and DeWeese}}]{zulkowski2012}
	\bibinfo{author}{\bibfnamefont{P.~R.} \bibnamefont{Zulkowski}},
	\bibinfo{author}{\bibfnamefont{D.~A.} \bibnamefont{Sivak}},
	\bibinfo{author}{\bibfnamefont{G.~E.} \bibnamefont{Crooks}},
	\bibnamefont{and} \bibinfo{author}{\bibfnamefont{M.~R.}
		\bibnamefont{DeWeese}}, \bibinfo{journal}{Phys. Rev. E}
	\textbf{\bibinfo{volume}{86}}, \bibinfo{pages}{041148}
	(\bibinfo{year}{2012}).
	
	\bibitem[{\citenamefont{Bonan{\c{c}}a and Deffner}(2014)}]{bonancca2014}
	\bibinfo{author}{\bibfnamefont{M.~V.} \bibnamefont{Bonan{\c{c}}a}}
	\bibnamefont{and} \bibinfo{author}{\bibfnamefont{S.}~\bibnamefont{Deffner}},
	\bibinfo{journal}{J. Chem. Phys.} \textbf{\bibinfo{volume}{140}},
	\bibinfo{pages}{244119} (\bibinfo{year}{2014}).
	
	\bibitem[{\citenamefont{Zulkowski and
			DeWeese}(2015{\natexlab{a}})}]{zulkowski2015}
	\bibinfo{author}{\bibfnamefont{P.~R.} \bibnamefont{Zulkowski}}
	\bibnamefont{and} \bibinfo{author}{\bibfnamefont{M.~R.}
		\bibnamefont{DeWeese}}, \bibinfo{journal}{Phys. Rev. E}
	\textbf{\bibinfo{volume}{92}}, \bibinfo{pages}{032117}
	(\bibinfo{year}{2015}{\natexlab{a}}).
	
	\bibitem[{\citenamefont{Zulkowski and
			DeWeese}(2015{\natexlab{b}})}]{zulkowski2015Quantum}
	\bibinfo{author}{\bibfnamefont{P.~R.} \bibnamefont{Zulkowski}}
	\bibnamefont{and} \bibinfo{author}{\bibfnamefont{M.~R.}
		\bibnamefont{DeWeese}}, \bibinfo{journal}{Phys. Rev. E}
	\textbf{\bibinfo{volume}{92}}, \bibinfo{pages}{032113}
	(\bibinfo{year}{2015}{\natexlab{b}}).
	
	\bibitem[{\citenamefont{Large and Sivak}(2019)}]{Large2019}
	\bibinfo{author}{\bibfnamefont{S.~J.} \bibnamefont{Large}} \bibnamefont{and}
	\bibinfo{author}{\bibfnamefont{D.~A.} \bibnamefont{Sivak}},
	\bibinfo{journal}{J. Stat. Mech.: Theory Exp.}
	\textbf{\bibinfo{volume}{2019}}, \bibinfo{pages}{083212}
	(\bibinfo{year}{2019}).
	
	\bibitem[{\citenamefont{Lucero et~al.}(2019)\citenamefont{Lucero, Mehdizadeh,
			and Sivak}}]{Lucero2019}
	\bibinfo{author}{\bibfnamefont{J.~N.} \bibnamefont{Lucero}},
	\bibinfo{author}{\bibfnamefont{A.}~\bibnamefont{Mehdizadeh}},
	\bibnamefont{and} \bibinfo{author}{\bibfnamefont{D.~A.} \bibnamefont{Sivak}},
	\bibinfo{journal}{Phys. Rev. E} \textbf{\bibinfo{volume}{99}},
	\bibinfo{pages}{012119} (\bibinfo{year}{2019}).
	
	\bibitem[{\citenamefont{Rotskoff and Crooks}(2015)}]{Rotskoff2015}
	\bibinfo{author}{\bibfnamefont{G.~M.} \bibnamefont{Rotskoff}} \bibnamefont{and}
	\bibinfo{author}{\bibfnamefont{G.~E.} \bibnamefont{Crooks}},
	\bibinfo{journal}{Phys. Rev. E} \textbf{\bibinfo{volume}{92}},
	\bibinfo{pages}{060102} (\bibinfo{year}{2015}).
	
	\bibitem[{\citenamefont{Rotskoff et~al.}(2017)\citenamefont{Rotskoff, Crooks,
			and Vanden-Eijnden}}]{Rotskoff2017}
	\bibinfo{author}{\bibfnamefont{G.~M.} \bibnamefont{Rotskoff}},
	\bibinfo{author}{\bibfnamefont{G.~E.} \bibnamefont{Crooks}},
	\bibnamefont{and}
	\bibinfo{author}{\bibfnamefont{E.}~\bibnamefont{Vanden-Eijnden}},
	\bibinfo{journal}{Phys. Rev. E} \textbf{\bibinfo{volume}{95}},
	\bibinfo{pages}{012148} (\bibinfo{year}{2017}).
	
	\bibitem[{\citenamefont{Louwerse and Sivak}(2022)}]{louwerse2022}
	\bibinfo{author}{\bibfnamefont{M.~D.} \bibnamefont{Louwerse}} \bibnamefont{and}
	\bibinfo{author}{\bibfnamefont{D.~A.} \bibnamefont{Sivak}},
	\bibinfo{journal}{J. Chem. Phys} \textbf{\bibinfo{volume}{156}},
	\bibinfo{pages}{194108} (\bibinfo{year}{2022}).
	
	\bibitem[{\citenamefont{Frim and DeWeese}(2022{\natexlab{a}})}]{frim2022}
	\bibinfo{author}{\bibfnamefont{A.~G.} \bibnamefont{Frim}} \bibnamefont{and}
	\bibinfo{author}{\bibfnamefont{M.~R.} \bibnamefont{DeWeese}},
	\bibinfo{journal}{Phys. Rev. E} \textbf{\bibinfo{volume}{105}},
	\bibinfo{pages}{L052103} (\bibinfo{year}{2022}{\natexlab{a}}).
	
	\bibitem[{\citenamefont{Frim and DeWeese}(2022{\natexlab{b}})}]{frim2021}
	\bibinfo{author}{\bibfnamefont{A.~G.} \bibnamefont{Frim}} \bibnamefont{and}
	\bibinfo{author}{\bibfnamefont{M.~R.} \bibnamefont{DeWeese}},
	\bibinfo{journal}{Phys. Rev. Lett.} \textbf{\bibinfo{volume}{128}},
	\bibinfo{pages}{230601} (\bibinfo{year}{2022}{\natexlab{b}}).
	
	\bibitem[{\citenamefont{Tafoya et~al.}(2019)\citenamefont{Tafoya, Large, Liu,
			Bustamante, and Sivak}}]{Tafoya2019}
	\bibinfo{author}{\bibfnamefont{S.}~\bibnamefont{Tafoya}},
	\bibinfo{author}{\bibfnamefont{S.~J.} \bibnamefont{Large}},
	\bibinfo{author}{\bibfnamefont{S.}~\bibnamefont{Liu}},
	\bibinfo{author}{\bibfnamefont{C.}~\bibnamefont{Bustamante}},
	\bibnamefont{and} \bibinfo{author}{\bibfnamefont{D.~A.} \bibnamefont{Sivak}},
	\bibinfo{journal}{Proc. Natl. Acad. Sci. U.S.A}
	\textbf{\bibinfo{volume}{116}}, \bibinfo{pages}{5920} (\bibinfo{year}{2019}).
	
	\bibitem[{\citenamefont{Blaber and
			Sivak}(2020{\natexlab{b}})}]{Blaber2020Skewed}
	\bibinfo{author}{\bibfnamefont{S.}~\bibnamefont{Blaber}} \bibnamefont{and}
	\bibinfo{author}{\bibfnamefont{D.~A.} \bibnamefont{Sivak}},
	\bibinfo{journal}{J. Chem. Phys.} \textbf{\bibinfo{volume}{153}},
	\bibinfo{pages}{244119} (\bibinfo{year}{2020}{\natexlab{b}}).
	
	\bibitem[{\citenamefont{Blaber and Sivak}(2022{\natexlab{b}})}]{Blaber2022}
	\bibinfo{author}{\bibfnamefont{S.}~\bibnamefont{Blaber}} \bibnamefont{and}
	\bibinfo{author}{\bibfnamefont{D.~A.} \bibnamefont{Sivak}},
	\bibinfo{journal}{Europhys. Lett.} \textbf{\bibinfo{volume}{139}},
	\bibinfo{pages}{17001} (\bibinfo{year}{2022}{\natexlab{b}}).
	
	\bibitem[{\citenamefont{Holubec and Ryabov}(2021)}]{holubec2021}
	\bibinfo{author}{\bibfnamefont{V.}~\bibnamefont{Holubec}} \bibnamefont{and}
	\bibinfo{author}{\bibfnamefont{A.}~\bibnamefont{Ryabov}},
	\bibinfo{journal}{J. Phys. A Math. Theor.} \textbf{\bibinfo{volume}{55}},
	\bibinfo{pages}{013001} (\bibinfo{year}{2021}).
	
	\bibitem[{\citenamefont{Ma et~al.}(2018{\natexlab{a}})\citenamefont{Ma, Xu,
			Dong, and Sun}}]{ma2018}
	\bibinfo{author}{\bibfnamefont{Y.-H.} \bibnamefont{Ma}},
	\bibinfo{author}{\bibfnamefont{D.}~\bibnamefont{Xu}},
	\bibinfo{author}{\bibfnamefont{H.}~\bibnamefont{Dong}}, \bibnamefont{and}
	\bibinfo{author}{\bibfnamefont{C.-P.} \bibnamefont{Sun}},
	\bibinfo{journal}{Phys. Rev. E} \textbf{\bibinfo{volume}{98}},
	\bibinfo{pages}{022133} (\bibinfo{year}{2018}{\natexlab{a}}).
	
	\bibitem[{\citenamefont{Zhang}(2020)}]{zhang2020}
	\bibinfo{author}{\bibfnamefont{Y.}~\bibnamefont{Zhang}}, \bibinfo{journal}{J.
		Stat. Phys.} \textbf{\bibinfo{volume}{178}}, \bibinfo{pages}{1336}
	(\bibinfo{year}{2020}).
	
	\bibitem[{\citenamefont{Abiuso and Perarnau-Llobet}(2020)}]{abiuso2020}
	\bibinfo{author}{\bibfnamefont{P.}~\bibnamefont{Abiuso}} \bibnamefont{and}
	\bibinfo{author}{\bibfnamefont{M.}~\bibnamefont{Perarnau-Llobet}},
	\bibinfo{journal}{Phys. Rev. Lett.} \textbf{\bibinfo{volume}{124}},
	\bibinfo{pages}{110606} (\bibinfo{year}{2020}).
	
	\bibitem[{\citenamefont{Chen}(2022)}]{chen2022}
	\bibinfo{author}{\bibfnamefont{J.-F.} \bibnamefont{Chen}},
	\bibinfo{journal}{Phys. Rev. E} \textbf{\bibinfo{volume}{106}},
	\bibinfo{pages}{054108} (\bibinfo{year}{2022}).
	
	\bibitem[{\citenamefont{Curzon and Ahlborn}(1975)}]{curzon1975}
	\bibinfo{author}{\bibfnamefont{F.~L.} \bibnamefont{Curzon}} \bibnamefont{and}
	\bibinfo{author}{\bibfnamefont{B.}~\bibnamefont{Ahlborn}},
	\bibinfo{journal}{Am. J. Phys.} \textbf{\bibinfo{volume}{43}},
	\bibinfo{pages}{22} (\bibinfo{year}{1975}).
	
	\bibitem[{\citenamefont{Van~den Broeck}(2005)}]{van2005}
	\bibinfo{author}{\bibfnamefont{C.}~\bibnamefont{Van~den Broeck}},
	\bibinfo{journal}{Phys. Rev. Lett.} \textbf{\bibinfo{volume}{95}},
	\bibinfo{pages}{190602} (\bibinfo{year}{2005}).
	
	\bibitem[{\citenamefont{Schmiedl and Seifert}(2008)}]{schmiedl2008efficiency}
	\bibinfo{author}{\bibfnamefont{T.}~\bibnamefont{Schmiedl}} \bibnamefont{and}
	\bibinfo{author}{\bibfnamefont{U.}~\bibnamefont{Seifert}},
	\bibinfo{journal}{Europhys. Lett.} \textbf{\bibinfo{volume}{83}},
	\bibinfo{pages}{30005} (\bibinfo{year}{2008}).
	
	\bibitem[{\citenamefont{Esposito et~al.}(2009)\citenamefont{Esposito,
			Lindenberg, and Van~den Broeck}}]{esposito2009}
	\bibinfo{author}{\bibfnamefont{M.}~\bibnamefont{Esposito}},
	\bibinfo{author}{\bibfnamefont{K.}~\bibnamefont{Lindenberg}},
	\bibnamefont{and} \bibinfo{author}{\bibfnamefont{C.}~\bibnamefont{Van~den
			Broeck}}, \bibinfo{journal}{Phys. Rev. Lett.} \textbf{\bibinfo{volume}{102}},
	\bibinfo{pages}{130602} (\bibinfo{year}{2009}).
	
	\bibitem[{\citenamefont{Esposito
			et~al.}(2010{\natexlab{b}})\citenamefont{Esposito, Kawai, Lindenberg, and
			Van~den Broeck}}]{esposito2010efficiency}
	\bibinfo{author}{\bibfnamefont{M.}~\bibnamefont{Esposito}},
	\bibinfo{author}{\bibfnamefont{R.}~\bibnamefont{Kawai}},
	\bibinfo{author}{\bibfnamefont{K.}~\bibnamefont{Lindenberg}},
	\bibnamefont{and} \bibinfo{author}{\bibfnamefont{C.}~\bibnamefont{Van~den
			Broeck}}, \bibinfo{journal}{Phys. Rev. Lett.} \textbf{\bibinfo{volume}{105}},
	\bibinfo{pages}{150603} (\bibinfo{year}{2010}{\natexlab{b}}).
	
	\bibitem[{\citenamefont{Brandner et~al.}(2015)\citenamefont{Brandner, Saito,
			and Seifert}}]{brandner2015}
	\bibinfo{author}{\bibfnamefont{K.}~\bibnamefont{Brandner}},
	\bibinfo{author}{\bibfnamefont{K.}~\bibnamefont{Saito}}, \bibnamefont{and}
	\bibinfo{author}{\bibfnamefont{U.}~\bibnamefont{Seifert}},
	\bibinfo{journal}{Phys. Rev. X.} \textbf{\bibinfo{volume}{5}},
	\bibinfo{pages}{031019} (\bibinfo{year}{2015}).
	
	\bibitem[{\citenamefont{Proesmans et~al.}(2016)\citenamefont{Proesmans,
			Cleuren, and Van~den Broeck}}]{proesmans2016}
	\bibinfo{author}{\bibfnamefont{K.}~\bibnamefont{Proesmans}},
	\bibinfo{author}{\bibfnamefont{B.}~\bibnamefont{Cleuren}}, \bibnamefont{and}
	\bibinfo{author}{\bibfnamefont{C.}~\bibnamefont{Van~den Broeck}},
	\bibinfo{journal}{Phys. Rev. Lett.} \textbf{\bibinfo{volume}{116}},
	\bibinfo{pages}{220601} (\bibinfo{year}{2016}).
	
	\bibitem[{\citenamefont{Shiraishi et~al.}(2016)\citenamefont{Shiraishi, Saito,
			and Tasaki}}]{shiraishi2016}
	\bibinfo{author}{\bibfnamefont{N.}~\bibnamefont{Shiraishi}},
	\bibinfo{author}{\bibfnamefont{K.}~\bibnamefont{Saito}}, \bibnamefont{and}
	\bibinfo{author}{\bibfnamefont{H.}~\bibnamefont{Tasaki}},
	\bibinfo{journal}{Phys. Rev. Lett.} \textbf{\bibinfo{volume}{117}},
	\bibinfo{pages}{190601} (\bibinfo{year}{2016}).
	
	\bibitem[{\citenamefont{Ma et~al.}(2018{\natexlab{b}})\citenamefont{Ma, Xu,
			Dong, and Sun}}]{ma2018universal}
	\bibinfo{author}{\bibfnamefont{Y.-H.} \bibnamefont{Ma}},
	\bibinfo{author}{\bibfnamefont{D.}~\bibnamefont{Xu}},
	\bibinfo{author}{\bibfnamefont{H.}~\bibnamefont{Dong}}, \bibnamefont{and}
	\bibinfo{author}{\bibfnamefont{C.-P.} \bibnamefont{Sun}},
	\bibinfo{journal}{Phys. Rev. E} \textbf{\bibinfo{volume}{98}},
	\bibinfo{pages}{042112} (\bibinfo{year}{2018}{\natexlab{b}}).
	
	\bibitem[{\citenamefont{Ma et~al.}(2020)\citenamefont{Ma, Zhai, Chen, Sun, and
			Dong}}]{ma2020}
	\bibinfo{author}{\bibfnamefont{Y.-H.} \bibnamefont{Ma}},
	\bibinfo{author}{\bibfnamefont{R.-X.} \bibnamefont{Zhai}},
	\bibinfo{author}{\bibfnamefont{J.}~\bibnamefont{Chen}},
	\bibinfo{author}{\bibfnamefont{C.}~\bibnamefont{Sun}}, \bibnamefont{and}
	\bibinfo{author}{\bibfnamefont{H.}~\bibnamefont{Dong}},
	\bibinfo{journal}{Phys. Rev. Lett.} \textbf{\bibinfo{volume}{125}},
	\bibinfo{pages}{210601} (\bibinfo{year}{2020}).
	
	\bibitem[{\citenamefont{Miller and Mehboudi}(2020)}]{miller2020}
	\bibinfo{author}{\bibfnamefont{H.~J.} \bibnamefont{Miller}} \bibnamefont{and}
	\bibinfo{author}{\bibfnamefont{M.}~\bibnamefont{Mehboudi}},
	\bibinfo{journal}{Phys. Rev. Lett.} \textbf{\bibinfo{volume}{125}},
	\bibinfo{pages}{260602} (\bibinfo{year}{2020}).
	
	\bibitem[{\citenamefont{Brandner and Saito}(2020)}]{brandner2020}
	\bibinfo{author}{\bibfnamefont{K.}~\bibnamefont{Brandner}} \bibnamefont{and}
	\bibinfo{author}{\bibfnamefont{K.}~\bibnamefont{Saito}},
	\bibinfo{journal}{Phys. Rev. Lett.} \textbf{\bibinfo{volume}{124}},
	\bibinfo{pages}{040602} (\bibinfo{year}{2020}).
	
	\bibitem[{\citenamefont{Miangolarra et~al.}(2021)\citenamefont{Miangolarra, Fu,
			Taghvaei, Chen, and Georgiou}}]{miangolarra2021}
	\bibinfo{author}{\bibfnamefont{O.~M.} \bibnamefont{Miangolarra}},
	\bibinfo{author}{\bibfnamefont{R.}~\bibnamefont{Fu}},
	\bibinfo{author}{\bibfnamefont{A.}~\bibnamefont{Taghvaei}},
	\bibinfo{author}{\bibfnamefont{Y.}~\bibnamefont{Chen}}, \bibnamefont{and}
	\bibinfo{author}{\bibfnamefont{T.~T.} \bibnamefont{Georgiou}},
	\bibinfo{journal}{Phys. Rev. E} \textbf{\bibinfo{volume}{103}},
	\bibinfo{pages}{062103} (\bibinfo{year}{2021}).
	
	\bibitem[{\citenamefont{Watanabe and Minami}(2022)}]{watanabe2022}
	\bibinfo{author}{\bibfnamefont{G.}~\bibnamefont{Watanabe}} \bibnamefont{and}
	\bibinfo{author}{\bibfnamefont{Y.}~\bibnamefont{Minami}},
	\bibinfo{journal}{Phys. Rev. Res.} \textbf{\bibinfo{volume}{4}},
	\bibinfo{pages}{L012008} (\bibinfo{year}{2022}).
	
	\bibitem[{\citenamefont{Acconcia et~al.}(2015)\citenamefont{Acconcia,
			Bonan{\c{c}}a, and Deffner}}]{acconcia2015}
	\bibinfo{author}{\bibfnamefont{T.~V.} \bibnamefont{Acconcia}},
	\bibinfo{author}{\bibfnamefont{M.~V.} \bibnamefont{Bonan{\c{c}}a}},
	\bibnamefont{and} \bibinfo{author}{\bibfnamefont{S.}~\bibnamefont{Deffner}},
	\bibinfo{journal}{Phys. Rev. E} \textbf{\bibinfo{volume}{92}},
	\bibinfo{pages}{042148} (\bibinfo{year}{2015}).
	
	\bibitem[{\citenamefont{Scandi and Perarnau-Llobet}(2019)}]{scandi2019}
	\bibinfo{author}{\bibfnamefont{M.}~\bibnamefont{Scandi}} \bibnamefont{and}
	\bibinfo{author}{\bibfnamefont{M.}~\bibnamefont{Perarnau-Llobet}},
	\bibinfo{journal}{Quantum} \textbf{\bibinfo{volume}{3}}, \bibinfo{pages}{197}
	(\bibinfo{year}{2019}).
	
	\bibitem[{\citenamefont{Takahashi}(2017)}]{takahashi2017}
	\bibinfo{author}{\bibfnamefont{K.}~\bibnamefont{Takahashi}},
	\bibinfo{journal}{New J. Phys.} \textbf{\bibinfo{volume}{19}},
	\bibinfo{pages}{115007} (\bibinfo{year}{2017}).
	
	\bibitem[{\citenamefont{Gu{\'e}ry-Odelin
			et~al.}(2019)\citenamefont{Gu{\'e}ry-Odelin, Ruschhaupt, Kiely, Torrontegui,
			Mart{\'\i}nez-Garaot, and Muga}}]{guery2019}
	\bibinfo{author}{\bibfnamefont{D.}~\bibnamefont{Gu{\'e}ry-Odelin}},
	\bibinfo{author}{\bibfnamefont{A.}~\bibnamefont{Ruschhaupt}},
	\bibinfo{author}{\bibfnamefont{A.}~\bibnamefont{Kiely}},
	\bibinfo{author}{\bibfnamefont{E.}~\bibnamefont{Torrontegui}},
	\bibinfo{author}{\bibfnamefont{S.}~\bibnamefont{Mart{\'\i}nez-Garaot}},
	\bibnamefont{and} \bibinfo{author}{\bibfnamefont{J.~G.} \bibnamefont{Muga}},
	\bibinfo{journal}{Rev. Mod. Phys} \textbf{\bibinfo{volume}{91}},
	\bibinfo{pages}{045001} (\bibinfo{year}{2019}).
	
	\bibitem[{\citenamefont{Gu{\'e}ry-Odelin
			et~al.}(2022)\citenamefont{Gu{\'e}ry-Odelin, Jarzynski, Plata, Prados, and
			Trizac}}]{guery2022}
	\bibinfo{author}{\bibfnamefont{D.}~\bibnamefont{Gu{\'e}ry-Odelin}},
	\bibinfo{author}{\bibfnamefont{C.}~\bibnamefont{Jarzynski}},
	\bibinfo{author}{\bibfnamefont{C.~A.} \bibnamefont{Plata}},
	\bibinfo{author}{\bibfnamefont{A.}~\bibnamefont{Prados}}, \bibnamefont{and}
	\bibinfo{author}{\bibfnamefont{E.}~\bibnamefont{Trizac}},
	\bibinfo{journal}{Rep. Prog. Phys.}  (\bibinfo{year}{2022}).
	
	\bibitem[{\citenamefont{Kumar and Bechhoefer}(2018)}]{kumar2018}
	\bibinfo{author}{\bibfnamefont{A.}~\bibnamefont{Kumar}} \bibnamefont{and}
	\bibinfo{author}{\bibfnamefont{J.}~\bibnamefont{Bechhoefer}},
	\bibinfo{journal}{Appl. Phys. Lett.} \textbf{\bibinfo{volume}{113}},
	\bibinfo{pages}{183702} (\bibinfo{year}{2018}).
	
	\bibitem[{\citenamefont{Kumar and Bechhoefer}(2019)}]{kumar2019}
	\bibinfo{author}{\bibfnamefont{A.}~\bibnamefont{Kumar}} \bibnamefont{and}
	\bibinfo{author}{\bibfnamefont{J.}~\bibnamefont{Bechhoefer}},
	\bibinfo{journal}{Physics in Canada} \textbf{\bibinfo{volume}{75}}
	(\bibinfo{year}{2019}).
	
	\bibitem[{\citenamefont{Gavrilov and Bechhoefer}(2017)}]{gavrilov2017}
	\bibinfo{author}{\bibfnamefont{M.}~\bibnamefont{Gavrilov}} \bibnamefont{and}
	\bibinfo{author}{\bibfnamefont{J.}~\bibnamefont{Bechhoefer}},
	\bibinfo{journal}{Philos. Transact. A Math. Phys. Eng. Sci.}
	\textbf{\bibinfo{volume}{375}}, \bibinfo{pages}{20160217}
	(\bibinfo{year}{2017}).
	
	\bibitem[{\citenamefont{Jun et~al.}(2014)\citenamefont{Jun, Gavrilov, and
			Bechhoefer}}]{jun2014}
	\bibinfo{author}{\bibfnamefont{Y.}~\bibnamefont{Jun}},
	\bibinfo{author}{\bibfnamefont{M.}~\bibnamefont{Gavrilov}}, \bibnamefont{and}
	\bibinfo{author}{\bibfnamefont{J.}~\bibnamefont{Bechhoefer}},
	\bibinfo{journal}{Phys. Rev. Lett.} \textbf{\bibinfo{volume}{113}},
	\bibinfo{pages}{190601} (\bibinfo{year}{2014}).
	
	\bibitem[{\citenamefont{Gavrilov and Bechhoefer}(2016)}]{gavrilov2016}
	\bibinfo{author}{\bibfnamefont{M.}~\bibnamefont{Gavrilov}} \bibnamefont{and}
	\bibinfo{author}{\bibfnamefont{J.}~\bibnamefont{Bechhoefer}},
	\bibinfo{journal}{Phys. Rev. Lett.} \textbf{\bibinfo{volume}{117}},
	\bibinfo{pages}{200601} (\bibinfo{year}{2016}).
	
	\bibitem[{\citenamefont{Neupane et~al.}(2015)\citenamefont{Neupane, Manuel,
			Lambert, and Woodside}}]{neupane2015}
	\bibinfo{author}{\bibfnamefont{K.}~\bibnamefont{Neupane}},
	\bibinfo{author}{\bibfnamefont{A.~P.} \bibnamefont{Manuel}},
	\bibinfo{author}{\bibfnamefont{J.}~\bibnamefont{Lambert}}, \bibnamefont{and}
	\bibinfo{author}{\bibfnamefont{M.~T.} \bibnamefont{Woodside}},
	\bibinfo{journal}{J. Phys. Chem. Lett} \textbf{\bibinfo{volume}{6}},
	\bibinfo{pages}{1005} (\bibinfo{year}{2015}).
	
	\bibitem[{\citenamefont{Hong et~al.}(2016)\citenamefont{Hong, Lambson, Dhuey,
			and Bokor}}]{hong2016}
	\bibinfo{author}{\bibfnamefont{J.}~\bibnamefont{Hong}},
	\bibinfo{author}{\bibfnamefont{B.}~\bibnamefont{Lambson}},
	\bibinfo{author}{\bibfnamefont{S.}~\bibnamefont{Dhuey}}, \bibnamefont{and}
	\bibinfo{author}{\bibfnamefont{J.}~\bibnamefont{Bokor}},
	\bibinfo{journal}{Sci. Adv.} \textbf{\bibinfo{volume}{2}},
	\bibinfo{pages}{e1501492} (\bibinfo{year}{2016}).
	
	\bibitem[{\citenamefont{Gupta et~al.}(2022)\citenamefont{Gupta, Large, Toyabe,
			and Sivak}}]{gupta2022}
	\bibinfo{author}{\bibfnamefont{D.}~\bibnamefont{Gupta}},
	\bibinfo{author}{\bibfnamefont{S.~J.} \bibnamefont{Large}},
	\bibinfo{author}{\bibfnamefont{S.}~\bibnamefont{Toyabe}}, \bibnamefont{and}
	\bibinfo{author}{\bibfnamefont{D.~A.} \bibnamefont{Sivak}},
	\bibinfo{journal}{J. Phys. Chem. Lett.} \textbf{\bibinfo{volume}{13}},
	\bibinfo{pages}{11844} (\bibinfo{year}{2022}).
	
	\bibitem[{\citenamefont{Park et~al.}(2003)\citenamefont{Park, Khalili-Araghi,
			Tajkhorshid, and Schulten}}]{Park2003}
	\bibinfo{author}{\bibfnamefont{S.}~\bibnamefont{Park}},
	\bibinfo{author}{\bibfnamefont{F.}~\bibnamefont{Khalili-Araghi}},
	\bibinfo{author}{\bibfnamefont{E.}~\bibnamefont{Tajkhorshid}},
	\bibnamefont{and} \bibinfo{author}{\bibfnamefont{K.}~\bibnamefont{Schulten}},
	\bibinfo{journal}{J. Chem. Phys.} \textbf{\bibinfo{volume}{119}},
	\bibinfo{pages}{3559} (\bibinfo{year}{2003}).
	
	\bibitem[{\citenamefont{Park and Schulten}(2004)}]{Park2004}
	\bibinfo{author}{\bibfnamefont{S.}~\bibnamefont{Park}} \bibnamefont{and}
	\bibinfo{author}{\bibfnamefont{K.}~\bibnamefont{Schulten}},
	\bibinfo{journal}{J. Chem. Phys.} \textbf{\bibinfo{volume}{120}},
	\bibinfo{pages}{5946} (\bibinfo{year}{2004}).
	
	\bibitem[{\citenamefont{Dellago and Hummer}(2014)}]{Dellago2014}
	\bibinfo{author}{\bibfnamefont{C.}~\bibnamefont{Dellago}} \bibnamefont{and}
	\bibinfo{author}{\bibfnamefont{G.}~\bibnamefont{Hummer}},
	\bibinfo{journal}{Entropy} \textbf{\bibinfo{volume}{16}}, \bibinfo{pages}{41}
	(\bibinfo{year}{2014}).
	
	\bibitem[{\citenamefont{Ito}(2023)}]{ito2022}
	\bibinfo{author}{\bibfnamefont{S.}~\bibnamefont{Ito}}, \bibinfo{journal}{Info.
		Geo.} pp. \bibinfo{pages}{1--42} (\bibinfo{year}{2023}).
	
	\bibitem[{\citenamefont{Abreu and Seifert}(2011)}]{Abreu2011}
	\bibinfo{author}{\bibfnamefont{D.}~\bibnamefont{Abreu}} \bibnamefont{and}
	\bibinfo{author}{\bibfnamefont{U.}~\bibnamefont{Seifert}},
	\bibinfo{journal}{Europhys. Lett.} \textbf{\bibinfo{volume}{94}},
	\bibinfo{pages}{10001} (\bibinfo{year}{2011}).
	
	\bibitem[{\citenamefont{Olkin and Pukelsheim}(1982)}]{Olkin1982}
	\bibinfo{author}{\bibfnamefont{I.}~\bibnamefont{Olkin}} \bibnamefont{and}
	\bibinfo{author}{\bibfnamefont{F.}~\bibnamefont{Pukelsheim}},
	\bibinfo{journal}{Linear Algebr. Appl.} \textbf{\bibinfo{volume}{48}},
	\bibinfo{pages}{257} (\bibinfo{year}{1982}).
	
	\bibitem[{\citenamefont{Dechant and Sakurai}(2019)}]{dechant2019}
	\bibinfo{author}{\bibfnamefont{A.}~\bibnamefont{Dechant}} \bibnamefont{and}
	\bibinfo{author}{\bibfnamefont{Y.}~\bibnamefont{Sakurai}},
	\bibinfo{journal}{arXiv preprint arXiv:1912.08405}  (\bibinfo{year}{2019}).
	
	\bibitem[{\citenamefont{Wang and Oster}(2002)}]{wang2002}
	\bibinfo{author}{\bibfnamefont{H.}~\bibnamefont{Wang}} \bibnamefont{and}
	\bibinfo{author}{\bibfnamefont{G.}~\bibnamefont{Oster}},
	\bibinfo{journal}{Europhys. Lett.} \textbf{\bibinfo{volume}{57}},
	\bibinfo{pages}{134} (\bibinfo{year}{2002}).
	
	\bibitem[{\citenamefont{Sivak and Crooks}(2012)}]{OptimalPaths}
	\bibinfo{author}{\bibfnamefont{D.~A.} \bibnamefont{Sivak}} \bibnamefont{and}
	\bibinfo{author}{\bibfnamefont{G.~E.} \bibnamefont{Crooks}},
	\bibinfo{journal}{Phys. Rev. Lett.} \textbf{\bibinfo{volume}{108}},
	\bibinfo{pages}{190602} (\bibinfo{year}{2012}).
	
	\bibitem[{\citenamefont{Mandal and Jarzynski}(2016)}]{mandal2016}
	\bibinfo{author}{\bibfnamefont{D.}~\bibnamefont{Mandal}} \bibnamefont{and}
	\bibinfo{author}{\bibfnamefont{C.}~\bibnamefont{Jarzynski}},
	\bibinfo{journal}{J. Stat. Mech. Theory Exp.}
	\textbf{\bibinfo{volume}{2016}}, \bibinfo{pages}{063204}
	(\bibinfo{year}{2016}).
	
	\bibitem[{\citenamefont{Zulkowski et~al.}(2013)\citenamefont{Zulkowski, Sivak,
			and DeWeese}}]{zulkowski2013}
	\bibinfo{author}{\bibfnamefont{P.~R.} \bibnamefont{Zulkowski}},
	\bibinfo{author}{\bibfnamefont{D.~A.} \bibnamefont{Sivak}}, \bibnamefont{and}
	\bibinfo{author}{\bibfnamefont{M.~R.} \bibnamefont{DeWeese}},
	\bibinfo{journal}{PloS one} \textbf{\bibinfo{volume}{8}},
	\bibinfo{pages}{e82754} (\bibinfo{year}{2013}).
	
	\bibitem[{\citenamefont{Large et~al.}(2018)\citenamefont{Large, Chetrite, and
			Sivak}}]{large2018}
	\bibinfo{author}{\bibfnamefont{S.~J.} \bibnamefont{Large}},
	\bibinfo{author}{\bibfnamefont{R.}~\bibnamefont{Chetrite}}, \bibnamefont{and}
	\bibinfo{author}{\bibfnamefont{D.~A.} \bibnamefont{Sivak}},
	\bibinfo{journal}{Europhys. Lett.} \textbf{\bibinfo{volume}{124}},
	\bibinfo{pages}{20001} (\bibinfo{year}{2018}).
	
	\bibitem[{\citenamefont{Oono and Paniconi}(1998)}]{oono1998}
	\bibinfo{author}{\bibfnamefont{Y.}~\bibnamefont{Oono}} \bibnamefont{and}
	\bibinfo{author}{\bibfnamefont{M.}~\bibnamefont{Paniconi}},
	\bibinfo{journal}{Progress of Theoretical Physics Supplement}
	\textbf{\bibinfo{volume}{130}}, \bibinfo{pages}{29} (\bibinfo{year}{1998}).
	
	\bibitem[{\citenamefont{Hatano and Sasa}(2001)}]{hatano2001}
	\bibinfo{author}{\bibfnamefont{T.}~\bibnamefont{Hatano}} \bibnamefont{and}
	\bibinfo{author}{\bibfnamefont{S.-i.} \bibnamefont{Sasa}},
	\bibinfo{journal}{Phys. Rev. Lett.} \textbf{\bibinfo{volume}{86}},
	\bibinfo{pages}{3463} (\bibinfo{year}{2001}).
	
	\bibitem[{\citenamefont{Speck and Seifert}(2005)}]{speck2005}
	\bibinfo{author}{\bibfnamefont{T.}~\bibnamefont{Speck}} \bibnamefont{and}
	\bibinfo{author}{\bibfnamefont{U.}~\bibnamefont{Seifert}},
	\bibinfo{journal}{J. Phys. A Math. Gen.} \textbf{\bibinfo{volume}{38}},
	\bibinfo{pages}{L581} (\bibinfo{year}{2005}).
	
	\bibitem[{\citenamefont{Ge}(2009)}]{ge2009}
	\bibinfo{author}{\bibfnamefont{H.}~\bibnamefont{Ge}}, \bibinfo{journal}{Phys.
		Rev. E} \textbf{\bibinfo{volume}{80}}, \bibinfo{pages}{021137}
	(\bibinfo{year}{2009}).
	
	\bibitem[{\citenamefont{Ge and Qian}(2010)}]{ge2010}
	\bibinfo{author}{\bibfnamefont{H.}~\bibnamefont{Ge}} \bibnamefont{and}
	\bibinfo{author}{\bibfnamefont{H.}~\bibnamefont{Qian}},
	\bibinfo{journal}{Phys. Rev. E} \textbf{\bibinfo{volume}{81}},
	\bibinfo{pages}{051133} (\bibinfo{year}{2010}).
	
	\bibitem[{\citenamefont{Bertini et~al.}(2012)\citenamefont{Bertini, Gabrielli,
			Jona-Lasinio, and Landim}}]{bertini2012}
	\bibinfo{author}{\bibfnamefont{L.}~\bibnamefont{Bertini}},
	\bibinfo{author}{\bibfnamefont{D.}~\bibnamefont{Gabrielli}},
	\bibinfo{author}{\bibfnamefont{G.}~\bibnamefont{Jona-Lasinio}},
	\bibnamefont{and} \bibinfo{author}{\bibfnamefont{C.}~\bibnamefont{Landim}},
	\bibinfo{journal}{J. Stat. Phys.} \textbf{\bibinfo{volume}{149}},
	\bibinfo{pages}{773} (\bibinfo{year}{2012}).
	
	\bibitem[{\citenamefont{Bertini et~al.}(2013)\citenamefont{Bertini, Gabrielli,
			Jona-Lasinio, and Landim}}]{bertini2013}
	\bibinfo{author}{\bibfnamefont{L.}~\bibnamefont{Bertini}},
	\bibinfo{author}{\bibfnamefont{D.}~\bibnamefont{Gabrielli}},
	\bibinfo{author}{\bibfnamefont{G.}~\bibnamefont{Jona-Lasinio}},
	\bibnamefont{and} \bibinfo{author}{\bibfnamefont{C.}~\bibnamefont{Landim}},
	\bibinfo{journal}{Phys. Rev. Lett.} \textbf{\bibinfo{volume}{110}},
	\bibinfo{pages}{020601} (\bibinfo{year}{2013}).
	
	\bibitem[{\citenamefont{Bertini et~al.}(2015)\citenamefont{Bertini, De~Sole,
			Gabrielli, Jona-Lasinio, and Landim}}]{bertini2015}
	\bibinfo{author}{\bibfnamefont{L.}~\bibnamefont{Bertini}},
	\bibinfo{author}{\bibfnamefont{A.}~\bibnamefont{De~Sole}},
	\bibinfo{author}{\bibfnamefont{D.}~\bibnamefont{Gabrielli}},
	\bibinfo{author}{\bibfnamefont{G.}~\bibnamefont{Jona-Lasinio}},
	\bibnamefont{and} \bibinfo{author}{\bibfnamefont{C.}~\bibnamefont{Landim}},
	\bibinfo{journal}{J. Stat. Mech. Theory Exp.}
	\textbf{\bibinfo{volume}{2015}}, \bibinfo{pages}{P10018}
	(\bibinfo{year}{2015}).
	
	\bibitem[{\citenamefont{Nulton et~al.}(1985)\citenamefont{Nulton, Salamon,
			Andresen, and Anmin}}]{nulton1985}
	\bibinfo{author}{\bibfnamefont{J.}~\bibnamefont{Nulton}},
	\bibinfo{author}{\bibfnamefont{P.}~\bibnamefont{Salamon}},
	\bibinfo{author}{\bibfnamefont{B.}~\bibnamefont{Andresen}}, \bibnamefont{and}
	\bibinfo{author}{\bibfnamefont{Q.}~\bibnamefont{Anmin}}, \bibinfo{journal}{J.
		Chem. Phys.} \textbf{\bibinfo{volume}{83}}, \bibinfo{pages}{334}
	(\bibinfo{year}{1985}).
	
	\bibitem[{\citenamefont{Bryant and Machta}(2020)}]{bryant2020}
	\bibinfo{author}{\bibfnamefont{S.~J.} \bibnamefont{Bryant}} \bibnamefont{and}
	\bibinfo{author}{\bibfnamefont{B.~B.} \bibnamefont{Machta}},
	\bibinfo{journal}{Proc. Natl. Acad. Sci. U.S.A.}
	\textbf{\bibinfo{volume}{117}}, \bibinfo{pages}{3478} (\bibinfo{year}{2020}).
	
	\bibitem[{\citenamefont{Machta}(2015)}]{machta2015}
	\bibinfo{author}{\bibfnamefont{B.~B.} \bibnamefont{Machta}},
	\bibinfo{journal}{Phys. Rev. Lett.} \textbf{\bibinfo{volume}{115}},
	\bibinfo{pages}{260603} (\bibinfo{year}{2015}).
	
	\bibitem[{\citenamefont{Gapsys et~al.}(2015)\citenamefont{Gapsys, Michielssens,
			Peters, de~Groot, and Leonov}}]{Gapsys2015}
	\bibinfo{author}{\bibfnamefont{V.}~\bibnamefont{Gapsys}},
	\bibinfo{author}{\bibfnamefont{S.}~\bibnamefont{Michielssens}},
	\bibinfo{author}{\bibfnamefont{J.~H.} \bibnamefont{Peters}},
	\bibinfo{author}{\bibfnamefont{B.~L.} \bibnamefont{de~Groot}},
	\bibnamefont{and} \bibinfo{author}{\bibfnamefont{H.}~\bibnamefont{Leonov}},
	in \emph{\bibinfo{booktitle}{Molecular Modeling of Proteins}}, edited by
	\bibinfo{editor}{\bibfnamefont{A.}~\bibnamefont{Kukol}}
	(\bibinfo{publisher}{Springer New York}, \bibinfo{address}{New York, NY},
	\bibinfo{year}{2015}), pp. \bibinfo{pages}{173--209}.
	
	\bibitem[{\citenamefont{Schindler et~al.}(2020)\citenamefont{Schindler,
			Baumann, Blum, B{\"o}se, Buchstaller, Burgdorf, Cappel, Chekler, Czodrowski,
			Dorsch et~al.}}]{Schindler2020}
	\bibinfo{author}{\bibfnamefont{C.~E.} \bibnamefont{Schindler}},
	\bibinfo{author}{\bibfnamefont{H.}~\bibnamefont{Baumann}},
	\bibinfo{author}{\bibfnamefont{A.}~\bibnamefont{Blum}},
	\bibinfo{author}{\bibfnamefont{D.}~\bibnamefont{B{\"o}se}},
	\bibinfo{author}{\bibfnamefont{H.-P.} \bibnamefont{Buchstaller}},
	\bibinfo{author}{\bibfnamefont{L.}~\bibnamefont{Burgdorf}},
	\bibinfo{author}{\bibfnamefont{D.}~\bibnamefont{Cappel}},
	\bibinfo{author}{\bibfnamefont{E.}~\bibnamefont{Chekler}},
	\bibinfo{author}{\bibfnamefont{P.}~\bibnamefont{Czodrowski}},
	\bibinfo{author}{\bibfnamefont{D.}~\bibnamefont{Dorsch}},
	\bibnamefont{et~al.}, \bibinfo{journal}{J. Chem. Inf. Model}
	(\bibinfo{year}{2020}).
	
	\bibitem[{\citenamefont{Kuhn et~al.}(2017)\citenamefont{Kuhn, Tich{\'y}, Wang,
			Robinson, Martin, Kuglstatter, Benz, Giroud, Schirmeister, Abel
			et~al.}}]{Kuhn2017}
	\bibinfo{author}{\bibfnamefont{B.}~\bibnamefont{Kuhn}},
	\bibinfo{author}{\bibfnamefont{M.}~\bibnamefont{Tich{\'y}}},
	\bibinfo{author}{\bibfnamefont{L.}~\bibnamefont{Wang}},
	\bibinfo{author}{\bibfnamefont{S.}~\bibnamefont{Robinson}},
	\bibinfo{author}{\bibfnamefont{R.~E.} \bibnamefont{Martin}},
	\bibinfo{author}{\bibfnamefont{A.}~\bibnamefont{Kuglstatter}},
	\bibinfo{author}{\bibfnamefont{J.}~\bibnamefont{Benz}},
	\bibinfo{author}{\bibfnamefont{M.}~\bibnamefont{Giroud}},
	\bibinfo{author}{\bibfnamefont{T.}~\bibnamefont{Schirmeister}},
	\bibinfo{author}{\bibfnamefont{R.}~\bibnamefont{Abel}}, \bibnamefont{et~al.},
	\bibinfo{journal}{J. Med. Chem} \textbf{\bibinfo{volume}{60}},
	\bibinfo{pages}{2485} (\bibinfo{year}{2017}).
	
	\bibitem[{\citenamefont{Ciordia et~al.}(2016)\citenamefont{Ciordia,
			P{\'e}rez-Benito, Delgado, Trabanco, and Tresadern}}]{Ciordia2016}
	\bibinfo{author}{\bibfnamefont{M.}~\bibnamefont{Ciordia}},
	\bibinfo{author}{\bibfnamefont{L.}~\bibnamefont{P{\'e}rez-Benito}},
	\bibinfo{author}{\bibfnamefont{F.}~\bibnamefont{Delgado}},
	\bibinfo{author}{\bibfnamefont{A.~A.} \bibnamefont{Trabanco}},
	\bibnamefont{and}
	\bibinfo{author}{\bibfnamefont{G.}~\bibnamefont{Tresadern}},
	\bibinfo{journal}{J. Chem. Inf. Model} \textbf{\bibinfo{volume}{56}},
	\bibinfo{pages}{1856} (\bibinfo{year}{2016}).
	
	\bibitem[{\citenamefont{Wang et~al.}(2015)\citenamefont{Wang, Wu, Deng, Kim,
			Pierce, Krilov, Lupyan, Robinson, Dahlgren, Greenwood et~al.}}]{Wang2015}
	\bibinfo{author}{\bibfnamefont{L.}~\bibnamefont{Wang}},
	\bibinfo{author}{\bibfnamefont{Y.}~\bibnamefont{Wu}},
	\bibinfo{author}{\bibfnamefont{Y.}~\bibnamefont{Deng}},
	\bibinfo{author}{\bibfnamefont{B.}~\bibnamefont{Kim}},
	\bibinfo{author}{\bibfnamefont{L.}~\bibnamefont{Pierce}},
	\bibinfo{author}{\bibfnamefont{G.}~\bibnamefont{Krilov}},
	\bibinfo{author}{\bibfnamefont{D.}~\bibnamefont{Lupyan}},
	\bibinfo{author}{\bibfnamefont{S.}~\bibnamefont{Robinson}},
	\bibinfo{author}{\bibfnamefont{M.~K.} \bibnamefont{Dahlgren}},
	\bibinfo{author}{\bibfnamefont{J.}~\bibnamefont{Greenwood}},
	\bibnamefont{et~al.}, \bibinfo{journal}{J. Am. Chem. Soc.}
	\textbf{\bibinfo{volume}{137}}, \bibinfo{pages}{2695} (\bibinfo{year}{2015}).
	
	\bibitem[{\citenamefont{Aldeghi et~al.}(2018)\citenamefont{Aldeghi, Gapsys, and
			de~Groot}}]{Aldeghi2018}
	\bibinfo{author}{\bibfnamefont{M.}~\bibnamefont{Aldeghi}},
	\bibinfo{author}{\bibfnamefont{V.}~\bibnamefont{Gapsys}}, \bibnamefont{and}
	\bibinfo{author}{\bibfnamefont{B.~L.} \bibnamefont{de~Groot}},
	\bibinfo{journal}{ACS Cent. Sci.} \textbf{\bibinfo{volume}{4}},
	\bibinfo{pages}{1708} (\bibinfo{year}{2018}).
	
	\bibitem[{\citenamefont{Chodera et~al.}(2011)\citenamefont{Chodera, Mobley,
			Shirts, Dixon, Branson, and Pande}}]{Chodera2011}
	\bibinfo{author}{\bibfnamefont{J.~D.} \bibnamefont{Chodera}},
	\bibinfo{author}{\bibfnamefont{D.~L.} \bibnamefont{Mobley}},
	\bibinfo{author}{\bibfnamefont{M.~R.} \bibnamefont{Shirts}},
	\bibinfo{author}{\bibfnamefont{R.~W.} \bibnamefont{Dixon}},
	\bibinfo{author}{\bibfnamefont{K.}~\bibnamefont{Branson}}, \bibnamefont{and}
	\bibinfo{author}{\bibfnamefont{V.~S.} \bibnamefont{Pande}},
	\bibinfo{journal}{Curr. Opin. Struc. Biol.} \textbf{\bibinfo{volume}{21}},
	\bibinfo{pages}{150} (\bibinfo{year}{2011}).
	
	\bibitem[{\citenamefont{Pohorille et~al.}(2010)\citenamefont{Pohorille,
			Jarzynski, and Chipot}}]{Pohorille2010}
	\bibinfo{author}{\bibfnamefont{A.}~\bibnamefont{Pohorille}},
	\bibinfo{author}{\bibfnamefont{C.}~\bibnamefont{Jarzynski}},
	\bibnamefont{and} \bibinfo{author}{\bibfnamefont{C.}~\bibnamefont{Chipot}},
	\bibinfo{journal}{J. Phys. Chem. B} \textbf{\bibinfo{volume}{114}},
	\bibinfo{pages}{10235} (\bibinfo{year}{2010}).
	
	\bibitem[{\citenamefont{Gore et~al.}(2003)\citenamefont{Gore, Ritort, and
			Bustamante}}]{Gore2003}
	\bibinfo{author}{\bibfnamefont{J.}~\bibnamefont{Gore}},
	\bibinfo{author}{\bibfnamefont{F.}~\bibnamefont{Ritort}}, \bibnamefont{and}
	\bibinfo{author}{\bibfnamefont{C.}~\bibnamefont{Bustamante}},
	\bibinfo{journal}{Proc. Natl. Acad. Sci. U.S.A}
	\textbf{\bibinfo{volume}{100}}, \bibinfo{pages}{12564}
	(\bibinfo{year}{2003}).
	
	\bibitem[{\citenamefont{Shenfeld et~al.}(2009)\citenamefont{Shenfeld, Xu,
			Eastwood, Dror, and Shaw}}]{Shenfeld2009}
	\bibinfo{author}{\bibfnamefont{D.~K.} \bibnamefont{Shenfeld}},
	\bibinfo{author}{\bibfnamefont{H.}~\bibnamefont{Xu}},
	\bibinfo{author}{\bibfnamefont{M.~P.} \bibnamefont{Eastwood}},
	\bibinfo{author}{\bibfnamefont{R.~O.} \bibnamefont{Dror}}, \bibnamefont{and}
	\bibinfo{author}{\bibfnamefont{D.~E.} \bibnamefont{Shaw}},
	\bibinfo{journal}{Phys. Rev. E} \textbf{\bibinfo{volume}{80}},
	\bibinfo{pages}{046705} (\bibinfo{year}{2009}).
	
	\bibitem[{\citenamefont{Minh}(2020)}]{Minh2019}
	\bibinfo{author}{\bibfnamefont{D.~D.~L.} \bibnamefont{Minh}},
	\bibinfo{journal}{J. Comput. Chem.} \textbf{\bibinfo{volume}{41}},
	\bibinfo{pages}{715} (\bibinfo{year}{2020}).
	
	\bibitem[{\citenamefont{Pham and Shirts}(2011)}]{Pham2011}
	\bibinfo{author}{\bibfnamefont{T.~T.} \bibnamefont{Pham}} \bibnamefont{and}
	\bibinfo{author}{\bibfnamefont{M.~R.} \bibnamefont{Shirts}},
	\bibinfo{journal}{J. Chem. Phys.} \textbf{\bibinfo{volume}{135}},
	\bibinfo{pages}{034114} (\bibinfo{year}{2011}).
	
	\bibitem[{\citenamefont{Pham and Shirts}(2012)}]{Pham2012}
	\bibinfo{author}{\bibfnamefont{T.~T.} \bibnamefont{Pham}} \bibnamefont{and}
	\bibinfo{author}{\bibfnamefont{M.~R.} \bibnamefont{Shirts}},
	\bibinfo{journal}{J. Chem. Phys.} \textbf{\bibinfo{volume}{136}},
	\bibinfo{pages}{124120} (\bibinfo{year}{2012}).
	
	\bibitem[{\citenamefont{Park and Im}(2014)}]{Park2014}
	\bibinfo{author}{\bibfnamefont{S.}~\bibnamefont{Park}} \bibnamefont{and}
	\bibinfo{author}{\bibfnamefont{W.}~\bibnamefont{Im}}, \bibinfo{journal}{J.
		Chem. Theory Comput.} \textbf{\bibinfo{volume}{10}}, \bibinfo{pages}{2719}
	(\bibinfo{year}{2014}).
	
	\bibitem[{\citenamefont{Jarzynski}(1997)}]{Jarzynski1997}
	\bibinfo{author}{\bibfnamefont{C.}~\bibnamefont{Jarzynski}},
	\bibinfo{journal}{Phys. Rev. Lett.} \textbf{\bibinfo{volume}{78}},
	\bibinfo{pages}{2690} (\bibinfo{year}{1997}).
	
	\bibitem[{\citenamefont{Bennett}(1976)}]{Bennett1976}
	\bibinfo{author}{\bibfnamefont{C.~H.} \bibnamefont{Bennett}},
	\bibinfo{journal}{J. Comput. Phys.} \textbf{\bibinfo{volume}{22}},
	\bibinfo{pages}{245} (\bibinfo{year}{1976}).
	
	\bibitem[{\citenamefont{Shirts et~al.}(2003)\citenamefont{Shirts, Bair, Hooker,
			and Pande}}]{Shirts2003}
	\bibinfo{author}{\bibfnamefont{M.~R.} \bibnamefont{Shirts}},
	\bibinfo{author}{\bibfnamefont{E.}~\bibnamefont{Bair}},
	\bibinfo{author}{\bibfnamefont{G.}~\bibnamefont{Hooker}}, \bibnamefont{and}
	\bibinfo{author}{\bibfnamefont{V.~S.} \bibnamefont{Pande}},
	\bibinfo{journal}{Phys. Rev. Lett.} \textbf{\bibinfo{volume}{91}},
	\bibinfo{pages}{140601} (\bibinfo{year}{2003}).
	
	\bibitem[{\citenamefont{Maragakis et~al.}(2006)\citenamefont{Maragakis,
			Spichty, and Karplus}}]{Maragakis2006}
	\bibinfo{author}{\bibfnamefont{P.}~\bibnamefont{Maragakis}},
	\bibinfo{author}{\bibfnamefont{M.}~\bibnamefont{Spichty}}, \bibnamefont{and}
	\bibinfo{author}{\bibfnamefont{M.}~\bibnamefont{Karplus}},
	\bibinfo{journal}{Phys. Rev. Lett.} \textbf{\bibinfo{volume}{96}},
	\bibinfo{pages}{100602} (\bibinfo{year}{2006}).
	
	\bibitem[{\citenamefont{Rolandi et~al.}(2022)\citenamefont{Rolandi,
			Perarnau-Llobet, and Miller}}]{rolandi2022}
	\bibinfo{author}{\bibfnamefont{A.}~\bibnamefont{Rolandi}},
	\bibinfo{author}{\bibfnamefont{M.}~\bibnamefont{Perarnau-Llobet}},
	\bibnamefont{and} \bibinfo{author}{\bibfnamefont{H.~J.}
		\bibnamefont{Miller}}, \bibinfo{journal}{arXiv preprint arXiv:2212.03927}
	(\bibinfo{year}{2022}).
	
	\bibitem[{\citenamefont{Best and Hummer}(2010)}]{best2010}
	\bibinfo{author}{\bibfnamefont{R.~B.} \bibnamefont{Best}} \bibnamefont{and}
	\bibinfo{author}{\bibfnamefont{G.}~\bibnamefont{Hummer}},
	\bibinfo{journal}{Proc. Natl. Acad. Sci. U.S.A.}
	\textbf{\bibinfo{volume}{107}}, \bibinfo{pages}{1088} (\bibinfo{year}{2010}).
	
	\bibitem[{\citenamefont{Hummer}(2005)}]{hummer2005}
	\bibinfo{author}{\bibfnamefont{G.}~\bibnamefont{Hummer}}, \bibinfo{journal}{New
		J. Phys.} \textbf{\bibinfo{volume}{7}}, \bibinfo{pages}{34}
	(\bibinfo{year}{2005}).
	
	\bibitem[{\citenamefont{Neupane et~al.}(2016)\citenamefont{Neupane, Manuel, and
			Woodside}}]{neupane2016}
	\bibinfo{author}{\bibfnamefont{K.}~\bibnamefont{Neupane}},
	\bibinfo{author}{\bibfnamefont{A.~P.} \bibnamefont{Manuel}},
	\bibnamefont{and} \bibinfo{author}{\bibfnamefont{M.~T.}
		\bibnamefont{Woodside}}, \bibinfo{journal}{Nat. Phys.}
	\textbf{\bibinfo{volume}{12}}, \bibinfo{pages}{700} (\bibinfo{year}{2016}).
	
	\bibitem[{\citenamefont{Foster et~al.}(2018)\citenamefont{Foster, Petrosyan,
			Pyo, Hoffmann, Wang, and Woodside}}]{foster2018}
	\bibinfo{author}{\bibfnamefont{D.~A.} \bibnamefont{Foster}},
	\bibinfo{author}{\bibfnamefont{R.}~\bibnamefont{Petrosyan}},
	\bibinfo{author}{\bibfnamefont{A.~G.} \bibnamefont{Pyo}},
	\bibinfo{author}{\bibfnamefont{A.}~\bibnamefont{Hoffmann}},
	\bibinfo{author}{\bibfnamefont{F.}~\bibnamefont{Wang}}, \bibnamefont{and}
	\bibinfo{author}{\bibfnamefont{M.~T.} \bibnamefont{Woodside}},
	\bibinfo{journal}{Biophys. J.} \textbf{\bibinfo{volume}{114}},
	\bibinfo{pages}{1657} (\bibinfo{year}{2018}).
	
	\bibitem[{\citenamefont{Chahine et~al.}(2007)\citenamefont{Chahine, Oliveira,
			Leite, and Wang}}]{chahine2007}
	\bibinfo{author}{\bibfnamefont{J.}~\bibnamefont{Chahine}},
	\bibinfo{author}{\bibfnamefont{R.~J.} \bibnamefont{Oliveira}},
	\bibinfo{author}{\bibfnamefont{V.~B.} \bibnamefont{Leite}}, \bibnamefont{and}
	\bibinfo{author}{\bibfnamefont{J.}~\bibnamefont{Wang}},
	\bibinfo{journal}{Proc. Natl. Acad. Sci. U.S.A.}
	\textbf{\bibinfo{volume}{104}}, \bibinfo{pages}{14646}
	(\bibinfo{year}{2007}).
	
	\bibitem[{\citenamefont{Bo et~al.}(2013)\citenamefont{Bo, Aurell, Eichhorn, and
			Celani}}]{bo2013}
	\bibinfo{author}{\bibfnamefont{S.}~\bibnamefont{Bo}},
	\bibinfo{author}{\bibfnamefont{E.}~\bibnamefont{Aurell}},
	\bibinfo{author}{\bibfnamefont{R.}~\bibnamefont{Eichhorn}}, \bibnamefont{and}
	\bibinfo{author}{\bibfnamefont{A.}~\bibnamefont{Celani}},
	\bibinfo{journal}{Europhys. Lett.} \textbf{\bibinfo{volume}{103}},
	\bibinfo{pages}{10010} (\bibinfo{year}{2013}).
	
	\bibitem[{\citenamefont{Proesmans}(2022)}]{proesmans2022}
	\bibinfo{author}{\bibfnamefont{K.}~\bibnamefont{Proesmans}},
	\bibinfo{journal}{arXiv preprint arXiv:2203.00428}  (\bibinfo{year}{2022}).
	
	\bibitem[{\citenamefont{Solon and Horowitz}(2018)}]{Solon2018}
	\bibinfo{author}{\bibfnamefont{A.~P.} \bibnamefont{Solon}} \bibnamefont{and}
	\bibinfo{author}{\bibfnamefont{J.~M.} \bibnamefont{Horowitz}},
	\bibinfo{journal}{Phys. Rev. Lett.} \textbf{\bibinfo{volume}{120}},
	\bibinfo{pages}{180605} (\bibinfo{year}{2018}).
	
	\bibitem[{\citenamefont{Ehrich and Sivak}(2023)}]{ehrich2022}
	\bibinfo{author}{\bibfnamefont{J.}~\bibnamefont{Ehrich}} \bibnamefont{and}
	\bibinfo{author}{\bibfnamefont{D.~A.} \bibnamefont{Sivak}},
	\bibinfo{journal}{Front. Phys.} \textbf{\bibinfo{volume}{11}}
	(\bibinfo{year}{2023}).
	
\end{thebibliography}
\end{document}